\documentclass[aps,prl,reprint,superscriptaddress,longbibliography,floatfix]{revtex4-2}

\usepackage{graphicx}
\usepackage{amsmath}
\usepackage{amssymb}
\usepackage{bm}
\usepackage{xcolor}
\usepackage{hyperref}

\newcommand{\baselineboxrow}[4]{%
\noalign{\hrule height \arrayrulewidth}%
\multicolumn{1}{|r}{#1} & #2 & #3 & \multicolumn{1}{r|}{#4} \\
\noalign{\hrule height \arrayrulewidth}%
}

\hypersetup{
  colorlinks=true,
  linkcolor=blue,
  citecolor=blue,
  urlcolor=blue
}

\begin{document}

\title{Hall Coefficient Sign Reversal Driven by Orbital-Selective Oxygen-Vacancy Scattering in Nickelate Films}

\author{Jian-Jian Miao}
\email{miaojianjian@quantumsc.cn}
\affiliation{Quantum Science Center of Guangdong-Hong Kong-Macao Greater Bay Area, Shenzhen 518045, China}

\author{Yue Liu}
\affiliation{Department of Physics and Guangdong Basic Research Center of Excellence for Quantum Science, Southern University of Science and Technology, Shenzhen 518055, China}

\author{Yue Zhao}
\affiliation{Department of Physics and Guangdong Basic Research Center of Excellence for Quantum Science, Southern University of Science and Technology, Shenzhen 518055, China}

\author{Yichen Hua}
\affiliation{Department of Physics and Guangdong Basic Research Center of Excellence for Quantum Science, Southern University of Science and Technology, Shenzhen 518055, China}
\affiliation{Quantum Science Center of Guangdong-Hong Kong-Macao Greater Bay Area, Shenzhen 518045, China}

\author{Changming Yue}
\email{yuecm@sustech.edu.cn}
\affiliation{Department of Physics and Guangdong Basic Research Center of Excellence for Quantum Science, Southern University of Science and Technology, Shenzhen 518055, China}
\affiliation{Quantum Science Center of Guangdong-Hong Kong-Macao Greater Bay Area, Shenzhen 518045, China}
\affiliation{Guangdong Provincial Key Laboratory of Advanced Thermoelectric Materials and Device Physics, Southern University of Science and Technology, Shenzhen 518055, China}

\author{Wei-Qiang Chen}
\email{chenwq@sustech.edu.cn}
\affiliation{Department of Physics and Guangdong Basic Research Center of Excellence for Quantum Science, Southern University of Science and Technology, Shenzhen 518055, China}
\affiliation{Quantum Science Center of Guangdong-Hong Kong-Macao Greater Bay Area, Shenzhen 518045, China}

\begin{abstract}
Hall measurements in superconducting bilayer nickelate films show sign reversals that cannot be explained by rigid-band electron doping alone. We combine a DFT+CDMFT-derived correlated multi-orbital quasiparticle model with a $T$-matrix treatment of oxygen-vacancy scattering in a semiclassical Boltzmann transport framework. We find that multiband compensation is insufficient by itself: in-plane vacancies selectively suppress the transport channel dominated by the $d_{x^2-y^2}$ orbital and drive $R_H$ through zero, whereas inner-apical vacancies make $R_H$ more negative. These results identify pocket-resolved and orbital-selective oxygen-vacancy scattering as the microscopic origin of the Hall coefficient sign reversal and provide a framework for oxygen-stoichiometry-dependent transport in nickelate films.
\end{abstract}

\maketitle

Nickelate superconductivity has rapidly emerged as an important frontier in correlated quantum materials, beginning with infinite-layer Nd$_{1-x}$Sr$_x$NiO$_2$ films~\cite{LiNature2019} and then extending to Ruddlesden-Popper bilayer La$_3$Ni$_2$O$_7$ under high pressure~\cite{SunNature2023}. More recently, ambient-pressure superconductivity in compressively strained bilayer films~\cite{KoNature2025,ZhouNature2025} has enabled structural, transport, optical, and spectroscopic probes that are difficult under pressure. Experiments now cover film and bulk transport~\cite{WangBulkNature2024,ZhangNatPhys2024,HaoNatMater2025,LiuTransportNatMater2025,WangDomePRL2026,LiPressureNatCommun2026,NieNature2026}, oxygen stoichiometry and defects~\cite{DongNature2024,DongNatMater2025,ShiSDWNatCommun2025}, ARPES, STM, XAS/RIXS, and optical spectroscopy~\cite{YangNatCommun2024,LiARPESNSR,Li3DARPES2026,ScienceAdvSTM,WangSTM2026,ChenRIXSNatCommun2024,MengUltrafastNatCommun2024}, and NMR/NQR, $\mu$SR, and neutron probes of density-wave order~\cite{ZhaoScienceBulletin2025,LuoNQRCPL2025,YashimaNQRJPSJ2025,ChenMuSRPRL2024,KhasanovNatPhys2025,PlokhikhNeutron2025,ChenNeutron2026}. In parallel, theory has addressed multiorbital electronic structures, correlations, and pairing from weak-coupling to strong-coupling approaches~\cite{LechermannPRB2023,FanXiangPRB2024,CaoYangPRB2024,GuLeHu2025,ZhangDagottoNatCommun2024,SakakibaraPRL2024,HeierSavrasovPRB2024,JiangWangQHPRL2025,XiaNatCommun2025,XiYuLiPRB2025,LuStrongPRL2024,YangOhZhangPRB2024,JiangWangZhangCPL2024,JiangHouKuPRL2024,LuoYaoNPJ2024,ZhangYouWengPRL2024,SchloemerCommunPhys2024,WangZhangJiangNSR2025,LangeStrongPRB2024,QuPRL2024,WangJiangZhangJin2026}; recent reviews summarize the broader landscape~\cite{NomuraAritaRPP2022,WangCPLReview2024,PuphalNatRevPhys2025,WangNSRReview2025,MiaoChenAPS2026}.

Hall transport has long been central to high-temperature superconductivity: in cuprates, $R_H$, the Hall number, and the Hall angle track carrier evolution~\cite{CuprateOngLSCO1987,CuprateHwang1994,CuprateAndo2004,CuprateBalakirev2003,CuprateBadoux2016}, reveal Fermi-surface reconstruction~\cite{CuprateDoironLeyraud2007,CuprateLeBoeuf2007,CuprateLaliberte2011}, and diagnose unconventional scattering~\cite{CuprateChien1991,CuprateAnderson1991,CuprateHarris1995}. In nickelate films, temperature-, field-, pressure-, and thickness-dependent measurements reveal positive, near-zero, and negative $R_H$ values and sign changes~\cite{KoNature2025,ZhouNature2025,HaoNatMater2025,LiuTransportNatMater2025,LiPressureNatCommun2026,ShiAdvMater2025}. Because film quality, doping, oxygen stoichiometry, disorder, and inhomogeneity vary among samples, a unified microscopic interpretation of $R_H$ remains unsettled. A recent half-dome experiment tracked $R_H$ across oxygen-rich, stoichiometric, and oxygen-deficient regimes and observed a Hall coefficient sign reversal upon introducing oxygen vacancies~\cite{LiuHalfDome}. Although such sign reversals have been broadly attributed to multiband effects, what selects the competing pocket and orbital transport channels has not been clarified microscopically.

In this Letter, we provide a microscopic theory for the Hall coefficient sign reversal in superconducting bilayer nickelate films. Starting from a correlated multi-orbital quasiparticle model, we incorporate a $T$-matrix treatment of elastic oxygen-vacancy scattering into a semiclassical Boltzmann transport framework. This framework separates rigid-band doping from vacancy-scattering-induced changes in pocket-resolved and orbital-selective transport relaxation times. We show that multiband compensation alone is insufficient: in-plane vacancies select the pocket and orbital channels by suppressing the $d_{x^2-y^2}$ orbital contribution more strongly than the $d_{z^2}$ contribution, thereby driving $R_H$ through zero. Our results provide a microscopic interpretation of the experimentally observed evolution of $R_H$ with oxygen stoichiometry and identify oxygen-vacancy scattering as a key control parameter for normal-state transport in bilayer nickelate films.

\begin{figure}[t]
\centering
\includegraphics[width=\columnwidth]{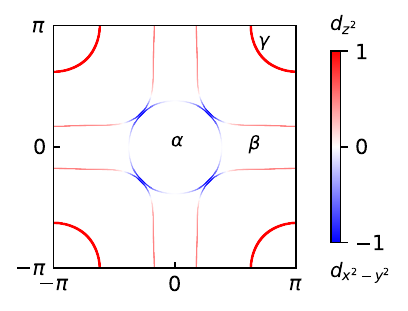}
\caption{Quasiparticle Fermi surface colored by the relative $d_{x^2-y^2}$ orbital weight $P^x=1-2W^x$.}
\label{fig:fs}
\end{figure}

\textit{Quasiparticle model.---}We start from the correlated multi-orbital model of the bilayer nickelate film obtained from DFT+CDMFT~\cite{YueNSR}, which reproduces the ARPES Fermiology of superconducting nickelate heterostructures~\cite{LiARPESNSR}. The low-energy electronic structure is captured by the four orbitals $(A_x,A_z,B_x,B_z)$, where $A/B$ labels the two NiO$_2$ layers and $x/z\equiv d_{x^2-y^2}/d_{z^2}$. Introducing the (anti-)bonding basis $(z_+,x_+,x_-,z_-)$, with $z_\pm=(A_z\pm B_z)/\sqrt{2}$ and $x_\pm=(A_x\pm B_x)/\sqrt{2}$, the quasiparticle Hamiltonian is $H_{\rm QP}=\sum_{{\bf k},\sigma}{\bf d}_{{\bf k}\sigma}^\dagger H_{\rm QP}({\bf k}){\bf d}_{{\bf k}\sigma}$, where ${\bf k}$ is the two-dimensional momentum, $\sigma$ is the spin index, and ${\bf d}_{{\bf k}\sigma}=(d_{{\bf k}z_+\sigma},d_{{\bf k}x_+\sigma},d_{{\bf k}x_-\sigma},d_{{\bf k}z_-\sigma})^T$. The quasiparticle Hamiltonian matrix reads
\begin{equation}
H_{\rm QP}({\bf k})
=
\hat Z^{1/2}
\left[
U^\dagger H_{\rm DFT}({\bf k})U
-\mu\hat I
+{\rm Re}\hat\Sigma(0)
\right]
\hat Z^{1/2},
\label{eq:hqp}
\end{equation}
where $H_{\rm DFT}({\bf k})$ is the DFT-derived tight-binding Hamiltonian in the layer-orbital basis, $U$ is the unitary matrix that transforms the layer-orbital basis to the (anti-)bonding basis, $\mu$ is the chemical potential, $\hat I$ is the identity matrix, ${\rm Re}\hat\Sigma(0)$ is the real part of the zero-frequency self-energy, and $\hat Z$ is the quasiparticle-weight matrix. In the DMFT calculation, both $\hat\Sigma$ and $\hat Z$ are diagonal in the (anti-)bonding basis; details are given in the \hyperref[app:qpf-parameters]{Supplemental Material}. The band structure is obtained by diagonalizing Eq.~(\ref{eq:hqp}), $H_{\rm QP}({\bf k})\phi_{m{\bf k}}=E_{m{\bf k}}\phi_{m{\bf k}}$, where $m$ is the band index, $E_{m{\bf k}}$ is the quasiparticle dispersion measured from the Fermi level, and $\phi_{m{\bf k}}$ is the eigenstate in the (anti-)bonding basis. The resulting quasiparticle Fermi surface, shown in Fig.~\ref{fig:fs}, consists of three pockets $\alpha$, $\beta$, and $\gamma$. We denote the orbital weight as $W_{m{\bf k}}^r=|\phi_{m{\bf k},r_+}|^2+|\phi_{m{\bf k},r_-}|^2$, $r=x,z$, and define the relative $d_{x^2-y^2}$ orbital weight as $P_{m{\bf k}}^x=1-2W_{m{\bf k}}^x$. Then the colored quasiparticle Fermi surface in Fig.~\ref{fig:fs} shows that the $\alpha$ and $\gamma$ pockets are predominantly derived from $d_{x^2-y^2}$ and $d_{z^2}$ orbitals, respectively, whereas the $\beta$ pocket has mixed orbital character.

\textit{Weak-field Hall coefficient.---}The key transport observable addressed here is the Hall coefficient of nickelate films. We evaluate it within the weak-field limit of the semiclassical Boltzmann transport framework. We label the three pocket contours by $C_u$, with $u=\alpha,\beta,\gamma$, and define the band velocity and mean-free-path vector as ${\bf v}^{u}_{\bf k}=\hbar^{-1}\nabla_{\bf k}E_{u{\bf k}}$ and ${\boldsymbol \ell}^{u}_{\bf k}=\tau^{u}_{\bf k}{\bf v}^{u}_{\bf k}$, respectively. Here $\tau^{u}_{\bf k}$ is the transport relaxation time. The longitudinal conductivity is written as a line integral over the pocket contours,
\begin{equation}
\sigma_{aa}
=\sum_u
\frac{g_s e^2}{(2\pi)^2\hbar}
\oint_{C_u}\frac{dk}{|{\bf v}^{u}_{\bf k}|}
\tau^{u}_{\bf k}\left({\bf v}^{u}_{\bf k}\cdot\hat{\bf e}_a\right)^2 ,
\label{eq:sigma-aa}
\end{equation}
with $a=x,y$ and spin degeneracy $g_s=2$. From Ong's geometrical interpretation, the weak-field Hall conductivity is related to the signed area $A_{\ell}^{u}$ swept out by ${\boldsymbol \ell}^{u}$ in mean-free-path space~\cite{Ong1991},
\begin{equation}
\begin{aligned}
\frac{\sigma_{xy}}{B}
&=\sum_u\frac{\sigma_{xy}^{u}}{B}
=\sum_u\frac{g_s e^3}{h^2}A_{\ell}^{u},\\
A_{\ell}^{u}
&=\frac{1}{2}\oint_{C_u}
(\ell_x^u d\ell_y^u-\ell_y^u d\ell_x^u).
\end{aligned}
\label{eq:ong}
\end{equation}
Owing to the separability of the pocket contours, we define the \textit{pocket-resolved} contributions to the Hall coefficient as
\begin{equation}
R_H=\sum_u R_H^u,\qquad
R_H^u=\frac{\sigma_{xy}^{u}/B}{\sigma_{xx}\sigma_{yy}},
\label{eq:pocket-rh-decomposition}
\end{equation}
where $R_H=\rho_{yx}/B\simeq\sigma_{xy}/(B\sigma_{xx}\sigma_{yy})$ has been used. The derivation of the Hall coefficient formula is given in the \hyperref[app:weak-field-hall]{Supplemental Material}. Within this geometrical picture, the sign of $R_H$ is determined by the total signed area $\sum_u A_{\ell}^{u}$, so that the sign reversal occurs when
\begin{equation}
\sum_u A_{\ell}^{u}=0.
\label{eq:sign-criterion}
\end{equation}
Since the three pockets carry distinct orbital characters, this compensation among pocket-resolved contributions also encodes the competition between orbital-selective contributions.

\textit{Vacancy-free limit.---}We roughly estimate the Hall coefficient without oxygen vacancies as the baseline for the subsequent calculations. In this vacancy-free limit, we assume a constant transport relaxation time, $\tau^u_{\bf k}=\tau_0$. Then, $\sigma_{aa}\propto\tau_0$ and $\sigma_{xy}\propto\tau_0^2$, so that $R_H\propto\sigma_{xy}/(\sigma_{xx}\sigma_{yy})$ is independent of $\tau_0$. Under this assumption, we obtain $R_H=-0.2817\times10^{-3}~{\rm cm^3/C}$, which has the same sign and order of magnitude as the experimental value reported for optimized, near-stoichiometric bilayer-nickelate films~\cite{LiuHalfDome}. This justifies using the quasiparticle Fermi surface as a reasonable starting point before introducing oxygen vacancies. Although the Hall coefficient is not directly orbital resolved, we introduce the pocket-averaged orbital weights and define the \textit{orbital-selective} contributions to the Hall coefficient as
\begin{equation}
\overline W_u^r=
\frac{\oint_{C_u}dk\,W_{u{\bf k}}^r/|{\bf v}_{\bf k}^u|}
{\oint_{C_u}dk/|{\bf v}_{\bf k}^u|},
\qquad
R_H^r=\sum_u\overline W_u^r R_H^u .
\end{equation}
The pocket-resolved and orbital-selective contributions to the vacancy-free Hall coefficient are listed in Table~\ref{tab:pocket-rh}. We note that the $\gamma$ pocket contributes only weakly to $R_H$. In Ong's geometrical interpretation, this weak contribution reflects the small mean-free-path contour formed by the weakly dispersive $\gamma$ pocket owing to its small Fermi velocities. Therefore, in the vacancy-free limit, the orbital selectivity primarily arises from the compensation between the $\alpha$ and $\beta$ pockets.

\begin{table}[t]
\caption{Pocket-resolved and orbital-selective contributions to the vacancy-free Hall coefficient. Values are in units of $10^{-3}~{\rm cm^3/C}$. The numerical calculation details are given in the \protect\hyperref[app:clean-hall-numerics]{Supplemental Material}.}
\label{tab:pocket-rh}
\begin{ruledtabular}
\begin{tabular}{lrlr}
Pocket & $R_H^u$ & Orbital & $R_H^r$ \\
\hline
$\alpha$ & $-1.2835$ & $d_{x^2-y^2}$ & $-0.5140$ \\
$\beta$ & $+0.9985$ & $d_{z^2}$ & $+0.2323$ \\
$\gamma$ & $+0.0033$ & & \\
\end{tabular}
\end{ruledtabular}
\end{table}

\begin{figure}[t]
\centering
\includegraphics[width=\columnwidth]{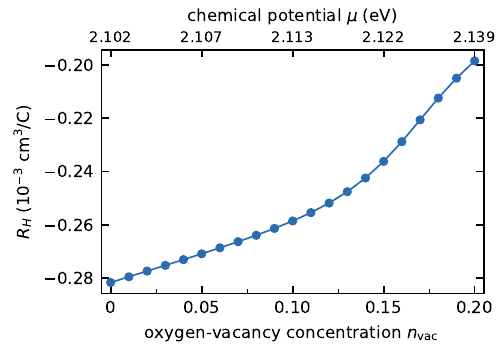}
\caption{Rigid-band $R_H$ as a function of $n_{\rm vac}$ and $\mu$.}
\label{fig:rigid-band-doping}
\end{figure}

\textit{Oxygen-vacancy doping effect.---}We now turn to the effect of oxygen vacancies on the Hall coefficient. Physically, their main role can be analyzed from two perspectives: the doping effect associated with oxygen removal and the impurity-scattering effect introduced by the oxygen-vacancy potential. To clarify the key ingredients for the Hall coefficient sign reversal, we consider these two effects separately. Theoretically, as a leading-order approximation, we model the doping effect within a rigid-band approximation by tuning the chemical potential of the quasiparticle band structure. Fig.~\ref{fig:rigid-band-doping} plots rigid-band $R_H$ as a function of the oxygen-vacancy concentration $n_{\rm vac}$ and the corresponding chemical potential $\mu$. Comparing the results in Tables~\ref{tab:pocket-rh} and \ref{tab:rigid-band-delta-minus}, the increase in the positive $\beta$-pocket contribution outweighs the enhanced negative $\alpha$-pocket contribution, making the rigid-band $R_H$ less negative. The orbital-selective contributions evolve in the same way as the $\alpha$- and $\beta$-pocket contributions. However, rigid-band $R_H$ remains negative over the range considered, suggesting that doping alone does not account for a Hall coefficient sign reversal.

\begin{table}[t]
\caption{Pocket-resolved and orbital-selective contributions to the rigid-band Hall coefficient at $n_{\rm vac}=0.1$. Values are in units of $10^{-3}~{\rm cm^3/C}$.}
\label{tab:rigid-band-delta-minus}
\begin{ruledtabular}
\begin{tabular}{lrlr}
Pocket & $R_H^u$ & Orbital & $R_H^r$ \\
\hline
$\alpha$ & $-1.3326$ & $d_{x^2-y^2}$ & $-0.5692$ \\
$\beta$ & $+1.0736$ & $d_{z^2}$ & $+0.3107$ \\
$\gamma$ & $+0.0005$ & & \\
\end{tabular}
\end{ruledtabular}
\end{table}

\textit{Orbital-selective scattering potential.---}To study the impurity-scattering effect, we construct the single-vacancy scattering potential from first-principles DFT calculations. In the dilute-vacancy limit, interactions between different oxygen vacancies are neglected, allowing inner-apical and in-plane vacancies to be treated separately. The vacancy-induced tight-binding parameter changes are obtained by applying the same DFT-Wannier workflow to the vacancy-free and relaxed single-vacancy structures; computational details are given in the \hyperref[app:vacancy-tb]{Supplemental Material}. Importantly, this procedure yields a microscopic tight-binding model of the local vacancy potential rather than phenomenological impurity parameters. For an inner-apical vacancy, the largest changes are the neighboring $d_{z^2}$ orbital onsite energy, $\Delta\epsilon_z=-1.12~{\rm eV}$, and the strong reduction of the interlayer $d_{z^2}$ hopping, $\Delta t_\perp^z=+0.45~{\rm eV}$. By contrast, an in-plane vacancy mainly affects the adjacent $d_{x^2-y^2}$ sector, with $\Delta\epsilon_x=-1.34~{\rm eV}$, together with sizable local changes in $d_{x^2-y^2}$ hopping and $d_{x^2-y^2}$-$d_{z^2}$ hybridization of order $0.2$--$0.3~{\rm eV}$. This strong orbital contrast originates from the vacancy-induced distortion of the oxygen octahedra, which modifies the local crystal-field environment, and provides the microscopic basis for orbital-selective oxygen-vacancy scattering.

\textit{Hall coefficient sign reversal.---}We next examine the impurity-scattering effects on the Hall coefficient from \textit{inner-apical} and \textit{in-plane} oxygen vacancies. Combining the $T$-matrix with the semiclassical Boltzmann transport framework, the scalar relaxation-time approximation gives
\begin{equation}
\begin{aligned}
\frac{1}{\tau_{m{\bf k}}}
&=
\frac{2\pi}{\hbar}n_{\rm vac}
\sum_{m'}
\int_{\rm BZ}\frac{d^2k'}{(2\pi)^2}
\left|T_{m'm}^R\right|^2
\left(
1-\hat{\bf v}_{m{\bf k}}\cdot\hat{\bf v}_{m'{\bf k}'}
\right) \\
&\quad\times
\delta\left(E_{m{\bf k}}-E_{m'{\bf k}'}\right),
\end{aligned}
\label{eq:tmatrix-rate-main}
\end{equation}
where $\tau_{m{\bf k}}$ is the transport relaxation time of the quasiparticle state $|m{\bf k}\rangle$ due to oxygen-vacancy elastic scattering, $\hat{\bf v}_{m{\bf k}}={\bf v}_{m{\bf k}}/|{\bf v}_{m{\bf k}}|$, and $T_{m'm}^R\equiv T_{m'm}^R({\bf k}',{\bf k};E_{m{\bf k}})$ is the retarded $T$-matrix in the band basis generated by the oxygen-vacancy scattering potential. Here $\delta$ denotes the Dirac delta function ensuring elastic energy conservation. The derivation is given in the \hyperref[app:tmatrix-transport]{Supplemental Material}. Since the total transport relaxation time is now band- and momentum-dependent, the Hall coefficient also depends on the vacancy-free transport relaxation time $\tau_0$. A calibration from the semiclassical Boltzmann transport framework to the longitudinal resistivity reported in Ref.~\cite{ZhouNature2025} gives a nominal $\tau_0\simeq5.8~{\rm fs}$. In consideration of other scattering channels such as inelastic scattering and incoherent transport, the actual vacancy-free transport relaxation time is usually larger by one to two orders of magnitude. Therefore, in the following calculations we take $\tau_0=50~{\rm fs}$. For a distribution with $a\%$ inner-apical and $b\%$ in-plane oxygen vacancies, the total transport relaxation time is obtained from Matthiessen's rule,
\begin{equation}
\frac{1}{\tau_{m{\bf k}}^{\rm tot}}
=
\frac{1}{\tau_0}
+
a\%\,
\frac{1}{\tau_{m{\bf k}}^{\rm ap}}
+
b\%\,
\frac{1}{\tau_{m{\bf k}}^{\rm pl}} ,
\label{eq:matthiessen-main}
\end{equation}
where $a+b=100$, and $\tau_{m{\bf k}}^{\rm ap}$ and $\tau_{m{\bf k}}^{\rm pl}$ are the transport relaxation times induced by inner-apical and in-plane oxygen vacancies, respectively. Figure~\ref{fig:hall-vacancy-scattering} shows three robust trends: inner-apical vacancies make $R_H$ more negative without a sign reversal, in-plane vacancies drive $R_H$ through zero, and mixed distributions reverse sign more readily as the in-plane-vacancy fraction increases. Thus, the multiband structure supplies competing Hall channels, but in-plane oxygen-vacancy scattering selects them and produces the sign reversal. The sign-change concentration $n_{\rm vac}^{\ast}$ defined by $R_H(n_{\rm vac}^{\ast})=0$ shifts only quantitatively when $\tau_0$ is varied from $10$ to $500~{\rm fs}$, and the saturation at large $n_{\rm vac}$ follows from $\sigma_{aa}\propto1/n_{\rm vac}$ and $\sigma_{xy}\propto1/n_{\rm vac}^2$ once vacancy scattering dominates Eq.~(\ref{eq:matthiessen-main}). Figure~\ref{fig:pocket-resolved-hall} resolves the mechanism for the mixed $20/80$ distribution: oxygen-vacancy scattering suppresses both $\alpha$ and $\beta$ pocket contributions, but the $\alpha$ contribution is reduced more strongly, driving the total Hall coefficient positive. The apparent increase of the $\gamma$ contribution originates mainly from the reduced longitudinal conductivities in the denominator of $R_H^\gamma$, rather than from an enhanced pocket-resolved Hall conductivity. In orbital space, the $d_{x^2-y^2}$ contribution decreases markedly while the $d_{z^2}$ contribution remains nearly unchanged, establishing the orbital-selective origin of the sign reversal.

\begin{center}
\refstepcounter{figure}\label{fig:hall-vacancy-scattering}
\centering
\includegraphics[width=\columnwidth]{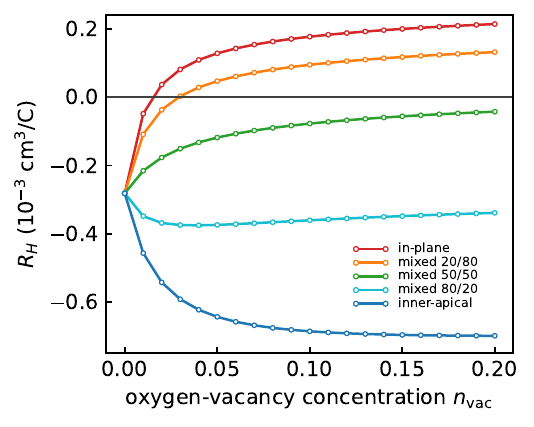}\\[1mm]
\small\textbf{FIG.~\thefigure.} $R_H$ as a function of $n_{\rm vac}$ for five oxygen-vacancy distributions, with $\tau_0=50~{\rm fs}$. A mixed distribution $a/b$ denotes $a\%$ inner-apical and $b\%$ in-plane oxygen vacancies.
\end{center}

\begin{center}
\refstepcounter{figure}\label{fig:pocket-resolved-hall}
\centering
\includegraphics[width=\columnwidth]{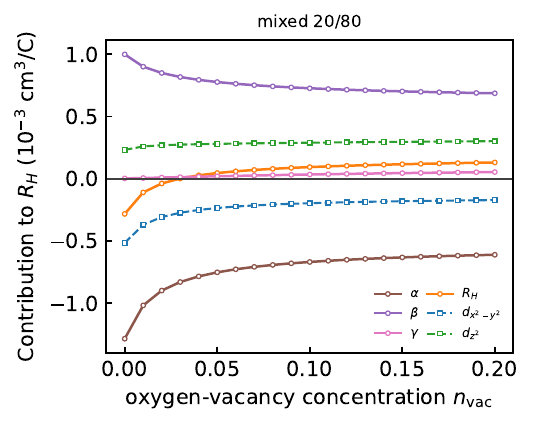}\\[1mm]
\small\textbf{FIG.~\thefigure.} Pocket-resolved (solid lines) and orbital-selective (dashed lines) contributions to $R_H$ for the mixed $20/80$ distribution, with $\tau_0=50~{\rm fs}$.
\end{center}

\textit{Discussion.---}Our results show that the multiband Fermi surface provides competing Hall channels, but the Hall coefficient sign reversal in bilayer nickelate films is selected by pocket-resolved and orbital-selective oxygen-vacancy scattering. In particular, in-plane oxygen vacancies selectively suppress the $d_{x^2-y^2}$ orbital contribution and thereby drive $R_H$ through zero, whereas inner-apical vacancies tend to make $R_H$ more negative. This distinction provides a microscopic interpretation for reconciling apparently diverse Hall responses among samples with different oxygen stoichiometry and vacancy distributions. Existing experiments on Sr-doped bilayer nickelate films indeed find that oxygen vacancies predominantly occupy in-plane rather than apical sites, further supporting our theoretical picture~\cite{HaoNatMater2025}.

In the study of high-temperature superconductors, many theories have been developed to explain the Hall effect and its anomalous behavior observed experimentally. These include marginal-Fermi-liquid and small-angle-scattering phenomenology~\cite{VarmaAbrahams2001,AbrahamsVarma2003}, strongly correlated theories based on doped Mott insulators or Hubbard/$t$-$J$ models~\cite{IoffeKalmeyerWiegmann1991,VebericPrelovsek2002}, and Fermi-surface reconstruction by stripe, density-wave, or antiferromagnetic order as a route to Hall coefficient sign changes~\cite{ChakravartyDDW2002,LinMillis2005,MillisNorman2007,EberleinSachdev2016}. In the nickelate films considered here, however, no pronounced resistive anomaly has been observed near the Hall sign reversal that would clearly indicate a corresponding ordered state. A more complete description of the temperature-, field-, pressure-, and thickness-dependent Hall response may require more sophisticated treatments, such as current vertex corrections in the presence of antiferromagnetic fluctuations~\cite{KankiKontani1999,KontaniKankiUeda1999}, which go beyond the semiclassical Boltzmann transport framework used in this Letter. Nevertheless, our results suggest that Hall measurements, when combined with structural probes of oxygen defects, can serve as a sensitive diagnostic of orbital-selective scattering in the normal state, and that oxygen vacancies should be treated as active scattering centers rather than merely as electron dopants in interpreting transport data of nickelate superconductors. Consistently, oxygen-content studies of bilayer nickelate films have found that, as oxygen defects increase and the system approaches the insulating regime, the resistance can be well fitted by Mott variable-range hopping behavior, also indicating the pronounced disorder effect induced by oxygen defects~\cite{WangElectronicStructuresSIT2025}.

\begin{acknowledgments}
Jian-Jian Miao acknowledges Heng Wang for discussion. Wei-Qiang Chen was supported by the National Key Research and Development Program of China (Grant No. 2024YFA1408101), the National Natural Science Foundation of China (Grant No. 12334002), the SUSTech-NUS Joint Research Program, the Science, Technology and Innovation Commission of Shenzhen Municipality (Grant No. ZDSYS20190902092905285), and the Center for Computational Science and Engineering at Southern University of Science and Technology. Changming Yue acknowledges support from the Guangdong Major Project of Basic Research (Grant No. 2025B0303000004), the Guangdong Provincial Quantum Science Strategic Initiative (Grant Nos. GDZX2401004 and GDZX2501001), and the National Natural Science Foundation of China (Grant No. 12474231). Jian-Jian Miao is supported by the National Natural Science Foundation of China (Grant No. 12404171) and the Guangdong Project (Grant No. 2024QN11X176).
\end{acknowledgments}

\clearpage

\bibliography{references}

@article{YueNSR,
  author = {Yue, Changming and Miao, Jian-Jian and Huang, Haoliang and Hua, Yichen and Li, Peng and Li, Yueying and Zhou, Guangdi and Lv, Wei and Yang, Qishuo and Sun, Hongyi and Sun, Yu-Jie and Lin, Junhao and Xue, Qi-Kun and Chen, Zhuoyu and Chen, Wei-Qiang},
  title = {Correlated electronic structures and unconventional superconductivity in bilayer nickelate heterostructures},
  journal = {National Science Review},
  volume = {12},
  number = {10},
  pages = {nwaf253},
  year = {2025},
  doi = {10.1093/nsr/nwaf253}
}

@article{LiNature2019,
  author = {Li, Danfeng and Lee, Kyuho and Wang, Bai Yang and Osada, Motoki and Crossley, Samuel and Lee, Hye Ryoung and Cui, Yi and Hikita, Yasuyuki and Hwang, Harold Y.},
  title = {Superconductivity in an infinite-layer nickelate},
  journal = {Nature},
  volume = {572},
  pages = {624--627},
  year = {2019},
  doi = {10.1038/s41586-019-1496-5}
}

@article{SunNature2023,
  author = {Sun, Hualei and Huo, Mengwu and Hu, Xunwu and Li, Jingyuan and Liu, Zengjia and Han, Yifeng and Tang, Lingyun and Mao, Zhongquan and Yang, Pengtao and Wang, Bosen and Cheng, Jinguang and Yao, Dao-Xin and Zhang, Guang-Ming and Wang, Meng},
  title = {Signatures of superconductivity near 80 {K} in a nickelate under high pressure},
  journal = {Nature},
  volume = {621},
  pages = {493--498},
  year = {2023},
  doi = {10.1038/s41586-023-06408-7}
}

@article{KoNature2025,
  author = {Ko, Eun Kyo and Yu, Yijun and Liu, Yidi and Bhatt, Lopa and Li, Jiarui and Thampy, Vivek and Kuo, Cheng-Tai and Wang, Bai Yang and Lee, Yonghun and Lee, Kyuho and Lee, Jun-Sik and Goodge, Berit H. and Muller, David A. and Hwang, Harold Y.},
  title = {Signatures of ambient pressure superconductivity in thin film {La$_3$Ni$_2$O$_7$}},
  journal = {Nature},
  volume = {638},
  number = {8052},
  pages = {935--940},
  year = {2025},
  doi = {10.1038/s41586-024-08525-3}
}

@article{ZhaoScienceBulletin2025,
  author = {Zhao, Dan and Zhou, Yanbing and Huo, Mengwu and Wang, Yu and Nie, Linpeng and Wang, Meng and Wu, Tao and Chen, Xianhui},
  title = {Pressure-enhanced spin-density-wave transition in double-layer nickelate La$_3$Ni$_2$O$_{7-\delta}$},
  journal = {Science Bulletin},
  volume = {70},
  pages = {1239--1245},
  year = {2025},
  doi = {10.1016/j.scib.2025.02.019}
}

@article{GuLeHu2025,
  author = {Gu, Y. H. and Le, C. C. and Yang, Z. S. and Wu, X. X. and Hu, J. P.},
  title = {Effective model and pairing tendency in bilayer Ni-based superconductor La$_3$Ni$_2$O$_7$},
  journal = {Physical Review B},
  volume = {111},
  pages = {174506},
  year = {2025},
  doi = {10.1103/PhysRevB.111.174506}
}

@article{LechermannPRB2023,
  author = {Lechermann, Frank and Gondolf, Jonas and B{\"o}tzel, Steffen and Eremin, Ilya M.},
  title = {Electronic correlations and superconducting instability in La$_3$Ni$_2$O$_7$ under high pressure},
  journal = {Physical Review B},
  volume = {108},
  pages = {L201121},
  year = {2023},
  doi = {10.1103/PhysRevB.108.L201121}
}

@article{FanXiangPRB2024,
  author = {Fan, Zhen and Zhang, Jian-Feng and Zhan, Bo and Lv, Dingshun and Jiang, Xing-Yu and Normand, Bruce and Xiang, Tao},
  title = {Superconductivity in nickelate and cuprate superconductors with strong bilayer coupling},
  journal = {Physical Review B},
  volume = {110},
  pages = {024514},
  year = {2024},
  doi = {10.1103/PhysRevB.110.024514}
}

@article{CaoYangPRB2024,
  author = {Cao, Yue and Yang, Yi-feng},
  title = {Flat bands promoted by Hund's rule coupling in the candidate double-layer high-temperature superconductor La$_3$Ni$_2$O$_7$ under high pressure},
  journal = {Physical Review B},
  volume = {109},
  pages = {L081105},
  year = {2024},
  doi = {10.1103/PhysRevB.109.L081105}
}

@article{SakakibaraPRL2024,
  author = {Sakakibara, H. and Kitamine, N. and Ochi, M. and Kuroki, K.},
  journal = {Physical Review Letters},
  volume = {132},
  pages = {106002},
  year = {2024},
  doi = {10.1103/PhysRevLett.132.106002}
}

@article{JiangWangQHPRL2025,
  author = {Jiang, K. Y. and Cao, Y. H. and Yang, Q. G. and Lu, H. Y. and Wang, Q. H.},
  title = {Theory of pressure dependence of superconductivity in bilayer nickelate La$_3$Ni$_2$O$_7$},
  journal = {Physical Review Letters},
  volume = {134},
  pages = {076001},
  year = {2025},
  doi = {10.1103/PhysRevLett.134.076001}
}

@article{XiaNatCommun2025,
  author = {Xia, C. and Liu, H. and Zhou, S. and Chen, H.},
  title = {Magnetic and electronic instabilities in pressurized La$_3$Ni$_2$O$_7$},
  journal = {Nature Communications},
  volume = {16},
  pages = {1054},
  year = {2025},
  doi = {10.1038/s41467-025-56455-5}
}

@article{LuStrongPRL2024,
  author = {Lu, C. and Pan, Z. and Yang, F. and Wu, C.},
  title = {Interlayer-coupling-driven high-temperature superconductivity in La$_3$Ni$_2$O$_7$ under pressure},
  journal = {Physical Review Letters},
  volume = {132},
  pages = {146002},
  year = {2024},
  doi = {10.1103/PhysRevLett.132.146002}
}

@article{ZhangYouWengPRL2024,
  author = {Zhang, J. X. and Zhang, H. K. and You, Y. Z. and Weng, Z. Y.},
  title = {Strong pairing originated from an emergent $Z_2$ Berry phase in La$_3$Ni$_2$O$_7$},
  journal = {Physical Review Letters},
  volume = {133},
  pages = {126501},
  year = {2024},
  doi = {10.1103/PhysRevLett.133.126501}
}

@article{WangZhangJiangNSR2025,
  author = {Wang, Z. and Zhang, H. J. and Jiang, K. and Zhang, F. C.},
  title = {Self-doped molecular Mott insulator for bilayer high-temperature superconducting La$_3$Ni$_2$O$_7$},
  journal = {National Science Review},
  volume = {12},
  pages = {nwaf353},
  year = {2025},
  doi = {10.1093/nsr/nwaf353}
}

@article{LangeStrongPRB2024,
  author = {Lange, H. and Homeier, L. and Demler, E. and Schollw{\"o}ck, U. and Bohrdt, A. and Grusdt, F.},
  title = {Feshbach resonance in a strongly repulsive ladder of mixed dimensionality: A possible scenario for bilayer nickelate superconductors},
  journal = {Physical Review B},
  volume = {109},
  pages = {045127},
  year = {2024},
  doi = {10.1103/PhysRevB.109.045127}
}

@article{ZhangDagottoNatCommun2024,
  author = {Zhang, Y. and Lin, L.-F. and Moreo, A. and Maier, T. A. and Dagotto, E.},
  title = {Structural phase transition, $s_{\pm}$-wave pairing, and magnetic stripe order in bilayered superconductor La$_3$Ni$_2$O$_7$ under pressure},
  journal = {Nature Communications},
  volume = {15},
  pages = {2470},
  year = {2024},
  doi = {10.1038/s41467-024-46622-1}
}

@article{HeierSavrasovPRB2024,
  author = {Heier, G. and Park, K. and Savrasov, S. Y.},
  title = {Competing $d_{xy}$ and $s_{\pm}$ pairing symmetries in superconducting La$_3$Ni$_2$O$_7$: LDA+FLEX calculations},
  journal = {Physical Review B},
  volume = {109},
  pages = {104508},
  year = {2024},
  doi = {10.1103/PhysRevB.109.104508}
}

@article{XiYuLiPRB2025,
  author = {Xi, W. and Yu, S.-L. and Li, J.-X.},
  title = {Transition from $s_{\pm}$-wave to $d_{x^2-y^2}$-wave superconductivity driven by interlayer interaction in the bilayer two-orbital model of La$_3$Ni$_2$O$_7$},
  journal = {Physical Review B},
  volume = {111},
  pages = {104505},
  year = {2025},
  doi = {10.1103/PhysRevB.111.104505}
}

@article{YangOhZhangPRB2024,
  author = {Yang, H. and Oh, H. and Zhang, Y.-H.},
  title = {Strong pairing from a small Fermi surface beyond weak coupling: Application to La$_3$Ni$_2$O$_7$},
  journal = {Physical Review B},
  volume = {110},
  pages = {104517},
  year = {2024},
  doi = {10.1103/PhysRevB.110.104517}
}

@article{JiangWangZhangCPL2024,
  author = {Jiang, Kun and Wang, Ziqiang and Zhang, Fu-Chun},
  title = {High temperature superconductivity in La$_3$Ni$_2$O$_7$},
  journal = {Chinese Physics Letters},
  volume = {41},
  pages = {017402},
  year = {2024},
  doi = {10.1088/0256-307X/41/1/017402}
}

@article{JiangHouKuPRL2024,
  author = {Jiang, R. and Hou, J. and Fan, Z. and Lang, Z.-J. and Ku, W.},
  title = {Pressure driven fractionalization of ionic spins results in cupratelike high-$T_c$ superconductivity in La$_3$Ni$_2$O$_7$},
  journal = {Physical Review Letters},
  volume = {132},
  pages = {126503},
  year = {2024},
  doi = {10.1103/PhysRevLett.132.126503}
}

@article{LuoYaoNPJ2024,
  author = {Luo, Z. and Lv, B. and Wang, M. and Wu, W. and Yao, D.-X.},
  title = {High-$T_c$ superconductivity in La$_3$Ni$_2$O$_7$ based on the bilayer two-orbital $t$-$J$ model},
  journal = {npj Quantum Materials},
  volume = {9},
  pages = {61},
  year = {2024},
  doi = {10.1038/s41535-024-00663-1}
}

@article{SchloemerCommunPhys2024,
  author = {Schl{\"o}mer, H. and Schollw{\"o}ck, U. and Grusdt, F. and Bohrdt, A.},
  title = {Superconductivity in the pressurized nickelate La$_3$Ni$_2$O$_7$ in the vicinity of a BEC-BCS crossover},
  journal = {Communications Physics},
  volume = {7},
  pages = {366},
  year = {2024},
  doi = {10.1038/s42005-024-01854-9}
}

@article{QuPRL2024,
  author = {Qu, X.-Z. and Qu, D.-W. and Chen, J. and Wu, C. and Yang, F. and Li, W. and Su, G.},
  title = {Bilayer $t$-$J$-$J_\perp$ model and magnetically mediated pairing in the pressurized nickelate La$_3$Ni$_2$O$_7$},
  journal = {Physical Review Letters},
  volume = {132},
  pages = {036502},
  year = {2024},
  doi = {10.1103/PhysRevLett.132.036502}
}

@article{WangJiangZhangJin2026,
  author = {Wang, Zhan and Jiang, Kun and Zhang, Fu-Chun and Jin, Hui-Ke},
  title = {Orbital-selective $d$-wave superconductivity in the two-band $t$-$J$ model for superconducting bilayer nickelates},
  journal = {arXiv},
  eprint = {2604.08319},
  year = {2026},
  doi = {10.48550/arXiv.2604.08319},
  url = {https://arxiv.org/abs/2604.08319}
}

@article{NomuraAritaRPP2022,
  author = {Nomura, Yusuke and Arita, Ryotaro},
  title = {Superconductivity in infinite-layer nickelates},
  journal = {Reports on Progress in Physics},
  volume = {85},
  number = {5},
  pages = {052501},
  year = {2022},
  doi = {10.1088/1361-6633/ac5a60}
}

@article{WangCPLReview2024,
  author = {Wang, Meng and Wen, Hai-Hu and Wu, Tao and Yao, Dao-Xin and Xiang, Tao},
  title = {Normal and superconducting properties of La$_3$Ni$_2$O$_7$},
  journal = {Chinese Physics Letters},
  volume = {41},
  number = {7},
  pages = {077402},
  year = {2024},
  doi = {10.1088/0256-307X/41/7/077402}
}

@article{WangNSRReview2025,
  author = {Wang, Yuxin and Jiang, Kun and Ying, Jianjun and Wu, Tao and Cheng, Jinguang and Hu, Jiangping and Chen, Xianhui},
  title = {Recent progress in nickelate superconductors},
  journal = {National Science Review},
  volume = {12},
  number = {10},
  pages = {nwaf373},
  year = {2025},
  doi = {10.1093/nsr/nwaf373}
}

@article{PuphalNatRevPhys2025,
  author = {Puphal, Pascal and Sch{\"a}fer, Thomas and Keimer, Bernhard and Hepting, Matthias},
  title = {Superconductivity in infinite-layer and Ruddlesden--Popper nickelates},
  journal = {Nature Reviews Physics},
  pages = {1--16},
  year = {2025},
  doi = {10.1038/s42254-025-00874-6}
}

@article{MiaoChenAPS2026,
  author = {Miao, Jianjian and Chen, Weiqiang},
  title = {Weak coupling theory of nickel-based 327 superconductors},
  journal = {Acta Physica Sinica},
  volume = {75},
  number = {7},
  pages = {070701},
  year = {2026},
  doi = {10.7498/aps.75.20252038}
}

@article{LiARPESNSR,
  author = {Li, Peng and Zhou, Guangdi and Lv, Wei and Li, Yueying and Yue, Changming and Huang, Haoliang and Xu, Lizhi and Shen, Jianchang and Miao, Yu and Song, Wenhua and Nie, Zihao and Chen, Yaqi and Wang, Heng and Chen, Weiqiang and Huang, Yaobo and Chen, Zhen-Hua and Qian, Tian and Lin, Junhao and He, Junfeng and Sun, Yu-Jie and Chen, Zhuoyu and Xue, Qi-Kun},
  title = {Angle-resolved photoemission spectroscopy of superconducting {(La,Pr)$_3$Ni$_2$O$_7$/SrLaAlO$_4$} heterostructures},
  journal = {National Science Review},
  volume = {12},
  number = {10},
  pages = {nwaf205},
  year = {2025},
  doi = {10.1093/nsr/nwaf205}
}

@article{YangNatCommun2024,
  author = {Yang, Jiangang and Sun, Hualei and Hu, Xunwu and Xie, Yuyang and Miao, Taimin and Luo, Hailan and Chen, Hao and Liang, Bo and Zhu, Wenpei and Qu, Gexing and Chen, Cui-Qun and Huo, Mengwu and Huang, Yaobo and Zhang, Shenjin and Zhang, Fengfeng and Yang, Feng and Wang, Zhimin and Peng, Qinjun and Mao, Hanqing and Liu, Guodong and Xu, Zuyan and Qian, Tian and Yao, Dao-Xin and Wang, Meng and Zhao, Lin and Zhou, X. J.},
  title = {Orbital-dependent electron correlation in double-layer nickelate La$_3$Ni$_2$O$_7$},
  journal = {Nature Communications},
  volume = {15},
  pages = {4373},
  year = {2024},
  doi = {10.1038/s41467-024-48701-7}
}

@article{Li3DARPES2026,
  author = {Li, Yueying and Xu, Lizhi and Lv, Wei and Nie, Zihao and Wang, Zechao and Miao, Yu and Shen, Jianchang and Zhou, Guangdi and Song, Wenhua and Wang, Heng and Huang, Haoliang and He, Junfeng and Jia, Jin-Feng and Li, Peng and Xue, Qi-Kun and Chen, Zhuoyu},
  title = {Three-dimensional electronic structures in superconducting Ruddlesden--Popper bilayer nickelate films},
  journal = {arXiv},
  eprint = {2604.08430},
  year = {2026}
}

@article{ScienceAdvSTM,
  author = {Fan, S. and Ou, M. and Scholten, M. and Li, Q. and Shang, Z. and Wang, Y. and Wen, H. H.},
  title = {Single-particle tunneling spectrum with a robust superconducting gap in {La$_2$PrNi$_2$O$_7$} thin films at ambient pressure},
  journal = {Science Advances},
  volume = {12},
  number = {24},
  pages = {eaeg2429},
  year = {2026},
  doi = {10.1126/sciadv.aeg2429}
}

@article{WangSTM2026,
  author = {Wang, Xinxin and Chen, Yaqi and Ding, Cui and Xu, Lizhi and Miao, Jian-Jian and Zhou, Guangdi and Chen, Zhuoyu and Sun, Yu-Jie and Jia, Jin-Feng and Xue, Qi-Kun},
  title = {Atomically resolved intrinsic superconducting gap in {(La, Pr)$_3$Ni$_2$O$_7$} films},
  journal = {arXiv},
  eprint = {2605.14806},
  year = {2026}
}

@article{LuoNQRCPL2025,
  author = {Luo, Jun and Feng, Jie and Wang, Gang and Wang, Ningning and Dou, Jie and Fang, Aifang and Yang, Jie and Cheng, Jinguang and Zheng, Guoqing and Zhou, Rui},
  title = {Microscopic evidence of charge- and spin-density waves in La$_3$Ni$_2$O$_{7-\delta}$ revealed by $^{139}$La-NQR},
  journal = {Chinese Physics Letters},
  volume = {42},
  pages = {067402},
  year = {2025},
  doi = {10.1088/0256-307X/42/6/067402}
}

@article{YashimaNQRJPSJ2025,
  author = {Yashima, Mitsuharu and Seto, Nina and Oshita, Yujiro and Kakoi, Masataka and Sakurai, Hiroya and Takano, Yoshihiko and Mukuda, Hidekazu},
  title = {Microscopic evidence for spin--spinless stripe order with reduced Ni moments within the $ab$ plane for bilayer nickelate La$_3$Ni$_2$O$_7$ probed by $^{139}$La-NQR},
  journal = {Journal of the Physical Society of Japan},
  volume = {94},
  pages = {054704},
  year = {2025},
  doi = {10.7566/JPSJ.94.054704}
}

@article{ChenMuSRPRL2024,
  author = {Chen, Kaiwen and Liu, Xiangqi and Jiao, Jiachen and Zou, Muyuan and Jiang, Chengyu and Li, Xin and Luo, Yixuan and Wu, Qiong and Zhang, Ningyuan and Guo, Yanfeng and Shu, Lei},
  title = {Evidence of spin density waves in La$_3$Ni$_2$O$_{7-\delta}$},
  journal = {Physical Review Letters},
  volume = {132},
  pages = {256503},
  year = {2024},
  doi = {10.1103/PhysRevLett.132.256503}
}

@article{KhasanovNatPhys2025,
  author = {Khasanov, Rustem and Hicken, Thomas J. and Gawryluk, Dariusz J. and Sazgari, Vahid and Plokhikh, Igor and Sorel, Lo{\"i}c Pierre and Bartkowiak, Marek and B{\"o}tzel, Steffen and Lechermann, Frank and Eremin, Ilya M. and Luetkens, Hubertus and Guguchia, Zurab},
  title = {Pressure-enhanced splitting of density wave transitions in La$_3$Ni$_2$O$_{7-\delta}$},
  journal = {Nature Physics},
  volume = {21},
  pages = {430--436},
  year = {2025},
  doi = {10.1038/s41567-024-02754-z}
}

@article{PlokhikhNeutron2025,
  author = {Plokhikh, Igor and Hicken, Thomas J. and Keller, Lukas and Pomjakushin, Vladimir and Moody, Samuel H. and Foury-Leylekian, Pascale and Krieger, Jonas J. and Luetkens, Hubertus and Guguchia, Zurab and Khasanov, Rustem and Gawryluk, Dariusz Jakub},
  title = {Unraveling spin density wave order in layered nickelates La$_3$Ni$_2$O$_7$ and La$_2$PrNi$_2$O$_7$ via neutron diffraction},
  journal = {arXiv},
  eprint = {2503.05287},
  year = {2025}
}

@article{ChenNeutron2026,
  author = {Chen, Lixing and Zhang, Enkang and Hao, Yiqing and Zhu, Yinghao and Cui, Bingkun and Abernathy, Douglas L. and Williams, Travis J. and Ikeda, Yoichi and Zhang, Hao and Liu, Feiyang and Wang, Wenbin and Wang, Qisi and Zhao, Jun},
  title = {Nature of magnetism in bilayer nickelate La$_3$Ni$_2$O$_7$ single crystals},
  journal = {arXiv},
  eprint = {2605.03448},
  year = {2026}
}

@article{ChenRIXSNatCommun2024,
  author = {Chen, Xiaoyang and Choi, Jaewon and Jiang, Zhicheng and Mei, Jiong and Jiang, Kun and Li, Jie and Agrestini, Stefano and Garcia-Fernandez, Mirian and Huang, Xing and Sun, Hualei and Shen, Dawei and Wang, Meng and Hu, Jiangping and Lu, Yi and Zhou, Ke-Jin and Feng, Donglai},
  title = {Electronic and magnetic excitations in La$_3$Ni$_2$O$_7$},
  journal = {Nature Communications},
  volume = {15},
  pages = {9597},
  year = {2024},
  doi = {10.1038/s41467-024-54104-6}
}

@article{DongNature2024,
  author = {Dong, Zehao and Huo, Mengwu and Li, Jie and Li, Jingyuan and Li, Pengcheng and Sun, Hualei and Lu, Yi and Wang, Meng and Wang, Yayu and Chen, Zhen},
  title = {Visualization of oxygen vacancies and self-doped ligand holes in La$_3$Ni$_2$O$_{7-\delta}$},
  journal = {Nature},
  volume = {630},
  pages = {847--852},
  year = {2024},
  doi = {10.1038/s41586-024-07482-1}
}

@article{DongNatMater2025,
  author = {Dong, Zehao and Wang, Gang and Wang, Ningning and Dong, Wen-Han and Gu, Lin and Xu, Yong and Cheng, Jinguang and Chen, Zhen and Wang, Yayu},
  title = {Interstitial oxygen order and its competition with superconductivity in La$_2$PrNi$_2$O$_{7+\delta}$},
  journal = {Nature Materials},
  volume = {24},
  pages = {1927--1934},
  year = {2025},
  doi = {10.1038/s41563-025-02351-2}
}

@article{ShiSDWNatCommun2025,
  author = {Shi, Mengzhu and Peng, Di and Li, Yikang and Yang, Shaohua and Xing, Zhenfang and Wang, Yuzhu and Fan, Kaibao and Li, Houpu and Wu, Rongqi and Ge, Binghui and Zeng, Zhidan and Zeng, Qiaoshi and Ying, Jianjun and Wu, Tao and Chen, Xianhui},
  title = {Spin density wave rather than tetragonal structure is prerequisite for superconductivity in La$_3$Ni$_2$O$_{7-\delta}$},
  journal = {Nature Communications},
  volume = {16},
  pages = {9141},
  year = {2025},
  doi = {10.1038/s41467-025-63701-x}
}

@article{MengUltrafastNatCommun2024,
  author = {Meng, Yanghao and Yang, Yi and Sun, Hualei and Zhang, Sasa and Luo, Jianlin and Chen, Liucheng and Ma, Xiaoli and Wang, Meng and Hong, Fang and Wang, Xinbo and Yu, Xiaohui},
  title = {Density-wave-like gap evolution in La$_3$Ni$_2$O$_7$ under high pressure revealed by ultrafast optical spectroscopy},
  journal = {Nature Communications},
  volume = {15},
  pages = {10408},
  year = {2024},
  doi = {10.1038/s41467-024-54771-5}
}

@article{NieNature2026,
  author = {Nie, Zihao and Li, Yueying and Lv, Wei and Xu, Lizhi and Jiang, Zhicheng and Fu, Peng and Zhou, Guangdi and Song, Wenhua and Chen, Yaqi and Wang, Heng and Huang, Haoliang and Lin, Junhao and Jia, Jin-Feng and Shen, Dawei and Li, Peng and Xue, Qi-Kun and Chen, Zhuoyu},
  title = {Superconductivity and electronic structures of nickelate thin film superstructures},
  journal = {Nature},
  volume = {652},
  pages = {628--634},
  year = {2026},
  doi = {10.1038/s41586-026-10352-7}
}

@article{WangDomePRL2026,
  author = {Wang, Meng and Sun, Hualei and Huo, Mengwu and Hu, Xunwu and Li, Jingyuan and Liu, Zengjia and Han, Yifeng and Tang, Lingyun and Mao, Zhongquan and Yang, Pengtao and Cheng, Jinguang and Yao, Dao-Xin and Zhang, Guang-Ming},
  title = {Superconducting dome in La$_{3-x}$Sr$_x$Ni$_2$O$_{7-\delta}$ thin films},
  journal = {Physical Review Letters},
  volume = {136},
  pages = {066002},
  year = {2026},
  doi = {10.1103/PhysRevLett.136.066002}
}

@article{WangBulkNature2024,
  author = {Wang, Ningning and Wang, Gang and Shen, Xiaoling and Hou, Jun and Luo, Jun and Ma, Xiaoping and Yang, Huaixin and Shi, Lifen and Dou, Jie and Feng, Jie and Yang, Jie and Shi, Yunqing and Ren, Zhian and Ma, Hanming and Yang, Pengtao and Liu, Ziyi and Liu, Yue and Zhang, Hua and Dong, Xiaoli and Wang, Yuxin and Jiang, Kun and Hu, Jiangping and Nagasaki, Shoko and Kitagawa, Kentaro and Calder, Stuart and Yan, Jiaqiang and Sun, Jianping and Wang, Bosen and Zhou, Rui and Uwatoko, Yoshiya and Cheng, Jinguang},
  title = {Bulk high-temperature superconductivity in pressurized tetragonal La$_2$PrNi$_2$O$_7$},
  journal = {Nature},
  volume = {634},
  pages = {579--584},
  year = {2024},
  doi = {10.1038/s41586-024-07996-8}
}

@article{ZhangNatPhys2024,
  author = {Zhang, Yanan and Su, Dajun and Huang, Yanen and Shan, Zhaoyang and Sun, Hualei and Huo, Mengwu and Ye, Kaixin and Zhang, Jiawen and Yang, Zihan and Xu, Yongkang and Su, Yi and Li, Rui and Smidman, Michael and Wang, Meng and Jiao, Lin and Yuan, Huiqiu},
  title = {High-temperature superconductivity with zero resistance and strange-metal behaviour in La$_3$Ni$_2$O$_{7-\delta}$},
  journal = {Nature Physics},
  volume = {20},
  pages = {1269--1273},
  year = {2024},
  doi = {10.1038/s41567-024-02515-y}
}

@article{Ong1991,
  author = {Ong, N. P.},
  title = {Geometric interpretation of the weak-field Hall conductivity in two-dimensional metals with arbitrary Fermi surface},
  journal = {Physical Review B},
  volume = {43},
  number = {1},
  pages = {193--201},
  year = {1991},
  doi = {10.1103/PhysRevB.43.193}
}

@article{CuprateOngLSCO1987,
  author = {Ong, N. P. and Wang, Z. Z. and Hagen, S. J. and Jing, T. W. and Clayhold, J. A. and Horvath, J.},
  title = {Hall effect of {La$_{2-x}$Sr$_x$CuO$_4$}: Implications for the electronic structure in the normal state},
  journal = {Physical Review B},
  volume = {35},
  number = {16},
  pages = {8807--8810},
  year = {1987},
  doi = {10.1103/PhysRevB.35.8807}
}

@article{CuprateChien1991,
  author = {Chien, T. R. and Wang, Z. Z. and Ong, N. P.},
  title = {Effect of {Zn} impurities on the normal-state Hall angle in single-crystal {YBa$_2$Cu$_{3-x}$Zn$_x$O$_{7-\delta}$}},
  journal = {Physical Review Letters},
  volume = {67},
  number = {15},
  pages = {2088--2091},
  year = {1991},
  doi = {10.1103/PhysRevLett.67.2088}
}

@article{CuprateAnderson1991,
  author = {Anderson, P. W.},
  title = {Hall effect in the two-dimensional {Luttinger} liquid},
  journal = {Physical Review Letters},
  volume = {67},
  number = {15},
  pages = {2092--2094},
  year = {1991},
  doi = {10.1103/PhysRevLett.67.2092}
}

@article{CuprateHwang1994,
  author = {Hwang, H. Y. and Batlogg, B. and Takagi, H. and Kao, H. L. and Kwo, J. and Cava, R. J. and Krajewski, J. J. and Peck, W. F.},
  title = {Scaling of the temperature dependent Hall effect in {La$_{2-x}$Sr$_x$CuO$_4$}},
  journal = {Physical Review Letters},
  volume = {72},
  number = {16},
  pages = {2636--2639},
  year = {1994},
  doi = {10.1103/PhysRevLett.72.2636}
}

@article{CuprateHarris1995,
  author = {Harris, J. M. and Yan, Y. F. and Matl, P. and Ong, N. P. and Anderson, P. W. and Kimura, T. and Kitazawa, K.},
  title = {Violation of {Kohler}'s rule in the normal-state magnetoresistance of {YBa$_2$Cu$_3$O$_{7-\delta}$} and {La$_{2-x}$Sr$_x$CuO$_4$}},
  journal = {Physical Review Letters},
  volume = {75},
  number = {7},
  pages = {1391--1394},
  year = {1995},
  doi = {10.1103/PhysRevLett.75.1391}
}

@article{CuprateBalakirev2003,
  author = {Balakirev, F. F. and Betts, J. B. and Migliori, A. and Ono, S. and Ando, Y. and Boebinger, G. S.},
  title = {Signature of optimal doping in Hall-effect measurements on a high-temperature superconductor},
  journal = {Nature},
  volume = {424},
  number = {6951},
  pages = {912--915},
  year = {2003},
  doi = {10.1038/nature01848}
}

@article{CuprateAndo2004,
  author = {Ando, Yoichi and Segawa, Kouji and Komiya, Seiki and Lavrov, A. N.},
  title = {Evolution of the Hall coefficient and the peculiar electronic structure of the cuprate superconductors},
  journal = {Physical Review Letters},
  volume = {92},
  number = {19},
  pages = {197001},
  year = {2004},
  doi = {10.1103/PhysRevLett.92.197001}
}

@article{CuprateDoironLeyraud2007,
  author = {Doiron-Leyraud, Nicolas and Proust, Cyril and LeBoeuf, David and Levallois, Julien and Bonnemaison, Jean-Baptiste and Liang, Ruixing and Bonn, D. A. and Hardy, W. N. and Taillefer, Louis},
  title = {Quantum oscillations and the Fermi surface in an underdoped high-$T_c$ superconductor},
  journal = {Nature},
  volume = {447},
  number = {7144},
  pages = {565--568},
  year = {2007},
  doi = {10.1038/nature05872}
}

@article{CuprateLeBoeuf2007,
  author = {LeBoeuf, D. and Doiron-Leyraud, N. and Levallois, J. and Daou, R. and Bonnemaison, J.-B. and Hussey, N. E. and Balicas, L. and Ramshaw, B. J. and Liang, R. and Bonn, D. A. and Hardy, W. N. and Adachi, S. and Proust, C. and Taillefer, L.},
  title = {Electron pockets in the Fermi surface of hole-doped high-$T_c$ superconductors},
  journal = {Nature},
  volume = {450},
  number = {7169},
  pages = {533--536},
  year = {2007},
  doi = {10.1038/nature06332}
}

@article{CuprateLaliberte2011,
  author = {Lalibert{\'e}, F. and Chang, J. and Doiron-Leyraud, N. and Hassinger, E. and Daou, R. and Rondeau, M. and Ramshaw, B. J. and Liang, R. and Bonn, D. A. and Hardy, W. N. and Pyon, S. and Takayama, T. and Takagi, H. and Sheikin, I. and Malone, L. and Proust, C. and Behnia, K. and Taillefer, L.},
  title = {Fermi-surface reconstruction by stripe order in cuprate superconductors},
  journal = {Nature Communications},
  volume = {2},
  pages = {432},
  year = {2011},
  doi = {10.1038/ncomms1440}
}

@article{CuprateBadoux2016,
  author = {Badoux, S. and Tabis, W. and Lalibert{\'e}, F. and Grissonnanche, G. and Vignolle, B. and Vignolles, D. and B{\'e}ard, J. and Bonn, D. A. and Hardy, W. N. and Liang, R. and Doiron-Leyraud, N. and Taillefer, L. and Proust, C.},
  title = {Change of carrier density at the pseudogap critical point of a cuprate superconductor},
  journal = {Nature},
  volume = {531},
  number = {7593},
  pages = {210--214},
  year = {2016},
  doi = {10.1038/nature16983}
}

@article{LiuHalfDome,
  author = {Liu, Yidi and Wang, Bai Yang and Li, Jiarui and Tarn, Yaoju and Bhatt, Lopa and Colletta, Michael and Wu, Yi-Ming and Kuo, Cheng-Tai and Lee, Jun-Sik and Goodge, Berit H. and Muller, David A. and Shen, Zhi-Xun and Raghu, Srinivas and Hwang, Harold Y. and Yu, Yijun},
  title = {A superconducting half-dome in bilayer nickelates},
  journal = {arXiv},
  eprint = {2603.12196},
  year = {2026}
}

@article{ZhouNature2025,
  author = {Zhou, Guangdi and Lv, Wei and Wang, Heng and Nie, Zihao and Chen, Yaqi and Li, Yueying and Huang, Haoliang and Chen, Wei-Qiang and Sun, Yu-Jie and Xue, Qi-Kun and Chen, Zhuoyu},
  title = {Ambient-pressure superconductivity onset above 40 {K} in {(La,Pr)$_3$Ni$_2$O$_7$} films},
  journal = {Nature},
  volume = {640},
  number = {8059},
  pages = {641},
  year = {2025},
  doi = {10.1038/s41586-025-08755-z}
}

@article{LiPressureNatCommun2026,
  author = {Li, Qing and Sun, Jianping and Boetzel, Steffen and Ou, Mengjun and Xiang, Zhe-Ning and Lechermann, Frank and Wang, Bosen and Wang, Yi and Zhang, Ying-Jie and Cheng, Jinguang and Eremin, Ilya M. and Wen, Hai-Hu},
  title = {Enhanced superconductivity in the compressively strained bilayer nickelate thin films by pressure},
  journal = {Nature Communications},
  volume = {17},
  pages = {3276},
  year = {2026},
  doi = {10.1038/s41467-026-69660-1}
}

@article{HaoNatMater2025,
  author = {Hao, Bo and Wang, Maosen and Sun, Wenjie and Yang, Yang and Mao, Zhangwen and Yan, Shengjun and Sun, Haoying and Zhang, Hongyi and Han, Lu and Gu, Zhengbin and Zhou, Jian and Ji, Dianxiang and Nie, Yuefeng},
  title = {Superconductivity in Sr-doped {La$_3$Ni$_2$O$_7$} thin films},
  journal = {Nature Materials},
  volume = {24},
  number = {11},
  pages = {1756--1762},
  year = {2025},
  doi = {10.1038/s41563-025-02327-2}
}

@article{LiuTransportNatMater2025,
  author = {Liu, Yidi and Ko, Eun Kyo and Tarn, Yaoju and Bhatt, Lopa and Li, Jiarui and Thampy, Vivek and Goodge, Berit H. and Muller, David A. and Raghu, Srinivas and Yu, Yijun and Hwang, Harold Y.},
  title = {Superconductivity and normal-state transport in compressively strained {La$_2$PrNi$_2$O$_7$} thin films},
  journal = {Nature Materials},
  volume = {24},
  number = {8},
  pages = {1221--1227},
  year = {2025},
  doi = {10.1038/s41563-025-02258-y}
}

@article{WangElectronicStructuresSIT2025,
  author = {Miao, Yu and Luan, Runqing and Chen, Yaqi and Ou, Zhipeng and Zhou, Guangdi and Shen, Jianchang and Wang, Heng and Huang, Haoliang and Wu, Xianfeng and Sun, Hongxu and Feng, Zikun and Yong, Xinru and Li, Yueying and Li, Peng and Xu, Lizhi and Lv, Wei and Nie, Zihao and Yue, Changming and Sun, Yu-Jie and Chen, Weiqiang and Yuan, Hongtao and Jia, Jin-Feng and Xue, Qi-Kun and Chen, Zhuoyu and He, Junfeng},
  title = {Electronic structures across superconductor-insulator transition in Ruddlesden-Popper bilayer nickelate films},
  journal = {arXiv},
  eprint = {2502.18068},
  year = {2025},
  doi = {10.48550/arXiv.2502.18068},
  url = {https://arxiv.org/abs/2502.18068}
}

@article{ShiAdvMater2025,
  author = {Shi, Yuexin and Song, Chenyao and Jia, Yingze and Wang, Yanzhi and Li, Qi and Chen, Ye and Yang, Yue and Fu, Junchi and Qin, Ming and Song, Dongsheng and Chen, Zhen and Yuan, Huiqiu and Xie, Yanwu and Zhang, Meng},
  title = {Critical Thickness and Long-Term Ambient Stability in Superconducting {LaPr$_2$Ni$_2$O$_7$} Films},
  journal = {Advanced Materials},
  volume = {38},
  number = {4},
  pages = {e10394},
  year = {2025},
  doi = {10.1002/adma.202510394}
}

@article{VarmaAbrahams2001,
  author = {Varma, C. M. and Abrahams, Elihu},
  title = {Effective Lorentz Force due to Small-Angle Impurity Scattering: Magnetotransport in High-$T_c$ Superconductors},
  journal = {Physical Review Letters},
  volume = {86},
  number = {20},
  pages = {4652--4655},
  year = {2001},
  doi = {10.1103/PhysRevLett.86.4652}
}

@article{AbrahamsVarma2003,
  author = {Abrahams, Elihu and Varma, C. M.},
  title = {Hall effect in the marginal Fermi liquid regime of high-$T_c$ superconductors},
  journal = {Physical Review B},
  volume = {68},
  number = {9},
  pages = {094502},
  year = {2003},
  doi = {10.1103/PhysRevB.68.094502}
}

@article{IoffeKalmeyerWiegmann1991,
  author = {Ioffe, L. B. and Kalmeyer, V. and Wiegmann, P. B.},
  title = {Hall coefficient of the doped Mott insulator: A signature of parity violation},
  journal = {Physical Review B},
  volume = {43},
  number = {1},
  pages = {1219--1222},
  year = {1991},
  doi = {10.1103/PhysRevB.43.1219}
}

@article{VebericPrelovsek2002,
  author = {Veberi{\v c}, D. and Prelov{\v s}ek, P.},
  title = {Temperature dependence of Hall response in doped antiferromagnets},
  journal = {Physical Review B},
  volume = {66},
  pages = {020408},
  year = {2002},
  doi = {10.1103/PhysRevB.66.020408}
}

@article{ChakravartyDDW2002,
  author = {Chakravarty, Sudip and Nayak, Chetan and Tewari, Sumanta and Yang, X.},
  title = {Sharp signature of a $d$-density wave quantum critical point in the Hall coefficient of the cuprates},
  journal = {Physical Review Letters},
  volume = {89},
  pages = {277003},
  year = {2002},
  doi = {10.1103/PhysRevLett.89.277003}
}

@article{LinMillis2005,
  author = {Lin, J. and Millis, A. J.},
  title = {Theory of low-temperature Hall effect in electron-doped cuprates},
  journal = {Physical Review B},
  volume = {72},
  pages = {214506},
  year = {2005},
  doi = {10.1103/PhysRevB.72.214506}
}

@article{MillisNorman2007,
  author = {Millis, A. J. and Norman, M. R.},
  title = {Antiphase stripe order as the origin of electron pockets observed in 1/8-hole-doped cuprates},
  journal = {Physical Review B},
  volume = {76},
  pages = {220503},
  year = {2007},
  doi = {10.1103/PhysRevB.76.220503}
}

@article{EberleinSachdev2016,
  author = {Eberlein, Andreas and Metzner, Walter and Sachdev, Subir and Yamase, Hiroyuki},
  title = {Fermi surface reconstruction and drop of the Hall number due to spiral antiferromagnetism in high-$T_c$ cuprates},
  journal = {Physical Review Letters},
  volume = {117},
  pages = {187001},
  year = {2016},
  doi = {10.1103/PhysRevLett.117.187001}
}

@article{KankiKontani1999,
  author = {Kanki, K. and Kontani, H.},
  title = {Theory of Hall effect and electrical transport in high-$T_c$ cuprates: Effects of antiferromagnetic spin fluctuations},
  journal = {Journal of the Physical Society of Japan},
  volume = {68},
  pages = {1614--1628},
  year = {1999},
  doi = {10.1143/JPSJ.68.1614},
  eprint = {cond-mat/9905428},
  archivePrefix = {arXiv},
  url = {https://arxiv.org/abs/cond-mat/9905428}
}

@article{KontaniKankiUeda1999,
  author = {Kontani, H. and Kanki, K. and Ueda, K.},
  title = {Hall effect and resistivity in high-$T_c$ superconductors: The conserving approximation},
  journal = {Physical Review B},
  volume = {59},
  pages = {14723--14739},
  year = {1999},
  doi = {10.1103/PhysRevB.59.14723}
}
\bibliographystyle{apsrev4-2}

\clearpage

\onecolumngrid

\appendix

\section{Supplemental Material A: Quasiparticle Hamiltonian and DFT+DMFT parameters}
\label{app:qpf-parameters}
\renewcommand{\theequation}{A.\arabic{equation}}
\renewcommand{\theHequation}{A.\arabic{equation}}
\setcounter{equation}{0}

We present the derivation of the quasiparticle Hamiltonian in the DFT+DMFT framework. In the layer-orbital basis,
\begin{equation}
\Psi_{\ell}
=
\left(A_x,A_z,B_x,B_z\right)^T ,
\label{eq:layer-spinor}
\end{equation}
$A_x$ and $A_z$ denote the bottom-layer Ni $d_{x^2-y^2}$ and $d_{z^2}$ orbitals, while $B_x$ and $B_z$ denote the corresponding top-layer orbitals. The (anti-)bonding basis is
\begin{equation}
\Psi_b
=
\left(z_+,x_+,x_-,z_-\right)^T ,
\qquad
z_\pm=\frac{A_z\pm B_z}{\sqrt{2}},
\quad
x_\pm=\frac{A_x\pm B_x}{\sqrt{2}} .
\label{eq:bonding-spinor}
\end{equation}
Writing $\Psi_\ell=U\Psi_b$, the unitary transformation matrix is
\begin{equation}
U
=
\frac{1}{\sqrt{2}}
\begin{pmatrix}
0 & 1 & 1 & 0 \\
1 & 0 & 0 & 1 \\
0 & 1 & -1 & 0 \\
1 & 0 & 0 & -1
\end{pmatrix}.
\label{eq:basis-u}
\end{equation}
Therefore the DFT tight-binding Hamiltonian in the (anti-)bonding basis is
\begin{equation}
\widetilde H_{\rm DFT}({\bf k})
=
U^\dagger H_{\rm DFT}({\bf k})U .
\label{eq:dft-rotation}
\end{equation}
In this basis the quasiparticle-weight matrix and the real part of the zero-frequency self-energy are
\begin{equation}
\hat Z
=
{\rm diag}
\left(
Z_{z_+},Z_{x_+},Z_{x_-},Z_{z_-}
\right),
\qquad
{\rm Re}\hat\Sigma(0)
=
{\rm diag}
\left(
{\rm Re}\Sigma_{z_+}(0),
{\rm Re}\Sigma_{x_+}(0),
{\rm Re}\Sigma_{x_-}(0),
{\rm Re}\Sigma_{z_-}(0)
\right).
\label{eq:z-sigma-diag}
\end{equation}
Keeping the linear-in-frequency part of the DMFT self-energy gives the quasiparticle self-energy
\begin{equation}
\hat\Sigma^{\rm QP}(\omega)
\simeq
{\rm Re}\hat\Sigma(0)
+
\left(\hat I-\hat Z^{-1}\right)(\omega+i0^+).
\label{eq:qp-self-energy}
\end{equation}
The quasiparticle Green's function is then
\begin{equation}
\begin{aligned}
\hat G_{\rm QP}({\bf k},\omega)
&=
\left[
(\omega+i0^++\mu)\hat I
-
\widetilde H_{\rm DFT}({\bf k})
-
\hat\Sigma^{\rm QP}(\omega)
\right]^{-1}\\
&=
\hat Z
\left[
(\omega+i0^+)\hat I
-
\left(
U^\dagger H_{\rm DFT}({\bf k})U
-\mu\hat I
+
{\rm Re}\hat\Sigma(0)
\right)\hat Z
\right]^{-1}\\
&=
\hat Z
\left[
(\omega+i0^+)\hat I
-
H_{\rm QP}({\bf k})
\right]^{-1}.
\end{aligned}
\label{eq:qp-green-derivation}
\end{equation}
Thus the single-particle matrix of the quasiparticle Hamiltonian in the symmetric form is
\begin{equation}
H_{\rm QP}({\bf k})
=
\hat Z^{1/2}
\left[
U^\dagger H_{\rm DFT}({\bf k})U
-\mu\hat I
+
{\rm Re}\hat\Sigma(0)
\right]
\hat Z^{1/2},
\label{eq:hqp-appendix}
\end{equation}
which is Eq.~(\ref{eq:hqp}) in the main text.

The DFT+DMFT parameters are collected in Tables~\ref{tab:tb-parameters} and~\ref{tab:qp-parameters}.

\begin{center}
\refstepcounter{table}\label{tab:tb-parameters}
\small\textbf{TABLE~\thetable.} DFT-derived tight-binding parameters. All parameters are given in eV.
\begin{ruledtabular}
\begin{tabular}{lrlr}
Parameter & \multicolumn{1}{c}{Value} & Parameter & \multicolumn{1}{c}{Value} \\
\hline
$\epsilon_x$ & $0.869804$ & $\epsilon_z$ & $0.350565$ \\
$t^{x}_{1}$ & $-0.466116$ & $t^{z}_{1}$ & $-0.126437$ \\
$t^{x}_{2}$ & $0.062314$ & $t^{z}_{2}$ & $-0.015764$ \\
$t^{x}_{4}$ & $-0.063894$ & $t^{z}_{4}$ & $-0.013821$ \\
$t^{x}_{5}$ & $-0.015354$ & $t^{z}_{5}$ & $-0.003271$ \\
$t^{xz}_{3}$ & $0.229225$ & $t^{xz}_{5}$ & $0.025474$ \\
$t^{x}_{\perp}$ & $0.001338$ & $t^{z}_{\perp}$ & $-0.438784$ \\
$t^{x}_{3}$ & $-0.000887$ & $t^{z}_{3}$ & $0.033221$ \\
$t^{xz}_{4}$ & $-0.031559$ & & \\
\end{tabular}
\end{ruledtabular}
\end{center}

\begin{center}
\refstepcounter{table}\label{tab:qp-parameters}
\small\textbf{TABLE~\thetable.} DMFT-derived quasiparticle parameters. The DMFT calculation uses $U=3.6~{\rm eV}$, $J=0.56~{\rm eV}$, and $n=1.3$ per Ni site.
\begin{ruledtabular}
\begin{tabular}{lc}
$\mu$ & $2.1017074585~{\rm eV}$ \\
$Z_{z_+}$ & $0.25118622$ \\
$Z_{x_+}$ & $0.54024784$ \\
$Z_{x_-}$ & $0.5418200890$ \\
$Z_{z_-}$ & $0.11676010$ \\
${\rm Re}\Sigma_{z_+}(0)$ & $1.9765884527~{\rm eV}$ \\
${\rm Re}\Sigma_{x_+}(0)$ & $2.2283682289~{\rm eV}$ \\
${\rm Re}\Sigma_{x_-}(0)$ & $2.2277489782~{\rm eV}$ \\
${\rm Re}\Sigma_{z_-}(0)$ & $2.4038664293~{\rm eV}$ \\
\end{tabular}
\end{ruledtabular}
\end{center}

To simplify the DFT+DMFT calculation, the parameters above were obtained in a half-unit-cell setup. In this reduced description, the $\alpha$ and $\beta$ pockets touch. In the full unit cell, however, the weak coupling between the two bilayers separates these pockets by opening a small gap. To make the $\alpha$ and $\beta$ pockets strictly separable in subsequent calculations, we add a small inter-pocket coupling
\begin{equation}
\begin{aligned}
H_{\alpha\beta}
&=
\Delta_{\alpha\beta}
\sum_{{\bf k},\sigma}
\left(
d^\dagger_{{\bf k}x_+\sigma}d_{{\bf k}x_-\sigma}
+
d^\dagger_{{\bf k}x_-\sigma}d_{{\bf k}x_+\sigma}
\right),
\end{aligned}
\label{eq:inter-pocket-coupling}
\end{equation}
where $d_{{\bf k}\zeta\sigma}$ annihilates an electron with spin $\sigma$ in the (anti-)bonding orbital $\zeta$. The quasiparticle Fermi-surface plot in Fig.~\ref{fig:fs} is evaluated on a $2048\times2048$ ${\bf k}$ mesh with the cutoff $|E_{m{\bf k}}|<0.5~{\rm meV}$ and $\Delta_{\alpha\beta}=15~{\rm meV}$. These choices balance resolving a visible small gap between the $\alpha$ and $\beta$ pockets with preserving the overall continuity of their pocket contours. We'll see that, provided the inter-pocket coupling is sufficiently small, its influence on the Hall coefficient is negligible.

\section{Supplemental Material B: Weak-field semiclassical Boltzmann transport and Hall coefficient formula}
\label{app:weak-field-hall}
\renewcommand{\theequation}{B.\arabic{equation}}
\renewcommand{\theHequation}{B.\arabic{equation}}
\setcounter{equation}{0}

We summarize the derivation of the Hall coefficient formula from the weak-field Boltzmann equation. For a quasiparticle state on pocket $u$, the steady-state Boltzmann equation in the relaxation-time approximation is
\begin{equation}
-\frac{e}{\hbar}
\left({\bf E}+{\bf v}^{u}_{\bf k}\times{\bf B}\right)
\cdot\nabla_{\bf k} f^{u}_{\bf k}
=
\frac{\delta f^{u}_{\bf k}}{\tau^u_{\bf k}},
\label{eq:boltzmann-rta}
\end{equation}
where $f^{u}_{\bf k}=f_0(E_{u{\bf k}})+\delta f^{u}_{\bf k}$ and ${\bf v}^{u}_{\bf k}=\hbar^{-1}\nabla_{\bf k}E_{u{\bf k}}$. The nonequilibrium distribution is parameterized by the vector mean-free path ${\boldsymbol \Lambda}^{u}_{\bf k}$,
\begin{equation}
\delta f^u_{\bf k}=
e\left(-\frac{\partial f_0}{\partial E}\right)
{\bf E}\cdot{\boldsymbol \Lambda}^{u}_{\bf k}.
\label{eq:delta-f}
\end{equation}
Substituting Eq.~(\ref{eq:delta-f}) into Eq.~(\ref{eq:boltzmann-rta}) and keeping terms linear in the electric field gives the standard recursion relation
\begin{equation}
{\boldsymbol \Lambda}^{u}_{\bf k}
=
\tau^u_{\bf k}
\left[
{\bf v}^{u}_{\bf k}
+\frac{e}{\hbar}({\bf v}^{u}_{\bf k}\times{\bf B})
\cdot\nabla_{\bf k}{\boldsymbol \Lambda}^{u}_{\bf k}
\right].
\label{eq:boltzmann-recursion}
\end{equation}
In weak magnetic field we expand
\begin{equation}
{\boldsymbol \Lambda}^{u}_{\bf k}
=
{\boldsymbol \Lambda}^{u,(0)}_{\bf k}
+{\boldsymbol \Lambda}^{u,(1)}_{\bf k}
+O(B^2).
\label{eq:lambda-expansion}
\end{equation}
The zeroth-order solution defines the mean-free-path vector used in the main text,
\begin{equation}
{\boldsymbol \Lambda}^{u,(0)}_{\bf k}
=
{\boldsymbol \ell}^{u}_{\bf k}
=
\tau^u_{\bf k}{\bf v}^{u}_{\bf k},
\label{eq:lambda-zero}
\end{equation}
while the first-order correction is
\begin{equation}
{\boldsymbol \Lambda}^{u,(1)}_{\bf k}
=
\tau^u_{\bf k}
\frac{e}{\hbar}
({\bf v}^{u}_{\bf k}\times{\bf B})
\cdot\nabla_{\bf k}{\boldsymbol \ell}^{u}_{\bf k}.
\label{eq:lambda-first}
\end{equation}

The charge current is obtained from
\begin{equation}
\begin{aligned}
j_a
&=g_s e\sum_u\int\frac{d^2k}{(2\pi)^2}
({\bf v}^{u}_{\bf k}\cdot\hat{\bf e}_a)\delta f^u_{\bf k}  \\
&\equiv \sum_b\sigma_{ab}E_b ,
\end{aligned}
\label{eq:current-appendix}
\end{equation}
Using Eq.~(\ref{eq:lambda-zero}) gives the longitudinal conductivity of pocket $u$,
\begin{equation}
\sigma_{aa}^{u}
=
g_s e^2
\int\frac{d^2k}{(2\pi)^2}
\left(-\frac{\partial f_0}{\partial E}\right)
\tau^u_{\bf k}({\bf v}^{u}_{\bf k}\cdot\hat{\bf e}_a)^2.
\label{eq:sigma-aa-bz}
\end{equation}
At low temperature, $-\partial f_0/\partial E$ restricts the integral to the Fermi contour. In the $T=0~{\rm K}$ limit,
\begin{equation}
\int d^2k\,\delta(E_{u{\bf k}})F({\bf k})
=
\oint_{C_u}
\frac{dk}{\hbar |{\bf v}^{u}_{\bf k}|}F({\bf k}),
\label{eq:fs-conversion}
\end{equation}
and therefore
\begin{equation}
\sigma_{aa}^{u}
=
\frac{g_s e^2}{(2\pi)^2\hbar}
\oint_{C_u}
\frac{dk}{|{\bf v}^{u}_{\bf k}|}
\tau^u_{\bf k}({\bf v}^{u}_{\bf k}\cdot\hat{\bf e}_a)^2 .
\label{eq:sigma-aa-appendix}
\end{equation}

The first-order correction in Eq.~(\ref{eq:lambda-first}) gives the weak-field Hall conductivity. Taking ${\bf B}=B\hat{\bf z}$ and collecting the coefficient of $E_y$ in $j_x$ yields
\begin{equation}
\frac{\sigma_{xy}^{u}}{B}
=\frac{g_s e^3}{(2\pi)^2\hbar^2}
\oint_{C_u}
\frac{dk}{|{\bf v}^{u}_{\bf k}|}
\ell_x^u
({\bf v}^{u}_{\bf k}\times\nabla_{\bf k})_z\ell_y^u .
\label{eq:sigma-xy-boltzmann}
\end{equation}
The contour coordinate obeys $d{\bf k}/ds=\hat{\bf z}\times{\bf v}^{u}_{\bf k}/|{\bf v}^{u}_{\bf k}|$. Hence Eq.~(\ref{eq:sigma-xy-boltzmann}) is the signed area swept out by ${\boldsymbol \ell}^{u}_{\bf k}$ in mean-free-path space,
\begin{equation}
\frac{\sigma_{xy}^{u}}{B}
=
\frac{g_s e^3}{h^2}A_\ell^u,
\qquad
A_\ell^u
=
\frac{1}{2}\oint_{C_u}
(\ell_x^u d\ell_y^u-\ell_y^u d\ell_x^u).
\label{eq:ong-appendix}
\end{equation}
This is Ong's geometric formula for pocket $u$, and the total Hall conductivity is obtained from $\sigma_{xy}/B=\sum_u\sigma_{xy}^{u}/B$. The area $A_\ell^u$ is signed by the orientation of the trajectory traced by ${\boldsymbol \ell}^{u}({\bf k})$; electron-like and hole-like pockets therefore contribute with opposite signs. In weak field,
\begin{equation}
R_H=\frac{\rho_{yx}}{B}
=\frac{1}{B}
\frac{\sigma_{xy}}{\sigma_{xx}\sigma_{yy}+\sigma_{xy}^2}
\simeq
\frac{\sigma_{xy}}{B\sigma_{xx}\sigma_{yy}},
\label{eq:rh-appendix}
\end{equation}
so the Hall coefficient changes sign precisely when the total signed Ong area $\sum_u A_\ell^u$ crosses zero.

\section{Supplemental Material C: Numerical calculation of the Hall coefficient without oxygen vacancies}
\label{app:clean-hall-numerics}
\renewcommand{\theequation}{C.\arabic{equation}}
\renewcommand{\theHequation}{C.\arabic{equation}}
\setcounter{equation}{0}

We describe the numerical procedure used to obtain the pocket-resolved and orbital-selective contributions to the vacancy-free Hall coefficient in Table~\ref{tab:pocket-rh}. Only the DFT+DMFT parameters in Tables~\ref{tab:tb-parameters} and~\ref{tab:qp-parameters} are used. The transport relaxation time is assumed to be a constant, $\tau^u_{\bf k}=\tau_0$, so the Hall coefficient is independent of $\tau_0$. Numerically, the Dirac delta function on the Fermi surface is represented by a Gaussian broadening,
\begin{equation}
\delta(E_{m{\bf q}})
\rightarrow
\delta_\eta(E_{m{\bf q}})
=
\frac{1}{\sqrt{\pi}\eta}
\exp\left[-\left(\frac{E_{m{\bf q}}}{\eta}\right)^2\right].
\label{eq:delta-eta-numerics}
\end{equation}
We use the physical wave vector ${\bf q}=(k_x/a,k_y/a)$ and ${\bf v}_{m{\bf q}}=\hbar^{-1}\nabla_{\bf q}E_{m{\bf q}}$. The derivatives entering the velocity and Hall integrand are evaluated by central differences with periodic boundary conditions on the Brillouin-zone mesh.

The three Fermi pockets are separated directly at the band level. After diagonalizing $H_{\rm QP}({\bf k})$ and sorting the eigenvalues, the bands crossing the Fermi level are assigned as $m=0$ for $\gamma$, $m=1$ for $\beta$, and $m=2$ for $\alpha$; the remaining band does not form a Fermi pocket and is excluded from the pocket-resolved sums. Figure~\ref{fig:pocket-assignment-numerics} shows the corresponding numerical grid points, selected only for visualization by $|E_{m{\bf k}}|<0.5~{\rm meV}$ with $\Delta_{\alpha\beta}=15~{\rm meV}$. The transport integrals below use the broadened Dirac delta function in Eq.~(\ref{eq:delta-eta-numerics}), not this visualization cutoff.

\begin{center}
\refstepcounter{figure}\label{fig:pocket-assignment-numerics}
\centering
\includegraphics[width=0.45\textwidth]{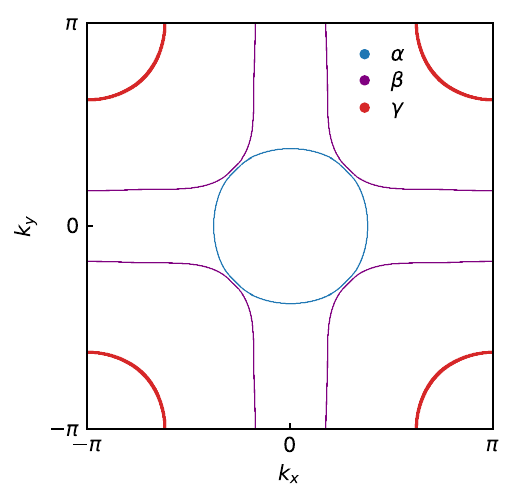}\\[1mm]
\small\textbf{FIG.~\thefigure.} Numerical pocket assignment on the $2048\times2048$ ${\bf k}$ mesh. Blue, magenta, and red points denote the $\alpha$, $\beta$, and $\gamma$ pockets, respectively.
\end{center}

With this pocket assignment, the pocket-resolved longitudinal conductivity is evaluated as
\begin{equation}
\frac{\sigma_{aa}^{u}}{\tau_0}
=
\frac{g_s e^2}{(2\pi)^2 c}
\sum_{m\in u}
\int_{\rm BZ} d^2q\,
\delta_\eta(E_{m{\bf q}})
\left({\bf v}_{m{\bf q}}\cdot\hat{\bf e}_a\right)^2 .
\label{eq:sigma-area-numerics}
\end{equation}
Here $g_s=2$ is the spin degeneracy and $c$ is the out-of-plane lattice constant. The corresponding weak-field Hall conductivity is evaluated as
\begin{equation}
\frac{1}{\tau_0^2}\frac{\sigma_{xy}^{u}}{B}
=
-\frac{g_s e^3}{(2\pi)^2\hbar c}
\sum_{m\in u}
\int_{\rm BZ} d^2q\,
\delta_\eta(E_{m{\bf q}})
\left({\bf v}_{m{\bf q}}\cdot\hat{\bf e}_x\right)
\left[
\left({\bf v}_{m{\bf q}}\times\nabla_{\bf q}\right)_z
\left({\bf v}_{m{\bf q}}\cdot\hat{\bf e}_y\right)
\right].
\label{eq:sigma-xy-area-numerics}
\end{equation}
The sign in Eq.~(\ref{eq:sigma-xy-area-numerics}) follows the convention $R_H=\rho_{yx}/B$. The pocket-resolved Hall contribution quoted in Table~\ref{tab:pocket-rh} is
\begin{equation}
R_H^u
=
\frac{\sigma_{xy}^{u}/B}{\sigma_{xx}\sigma_{yy}},
\label{eq:rh-pocket-numerics}
\end{equation}
where $\sigma_{aa}=\sum_u\sigma_{aa}^{u}$.

The orbital-selective contribution is obtained by assigning the pocket-resolved Hall contribution according to the pocket-averaged orbital character. We define the wave-function projection factors
\begin{equation}
{\cal P}^{x}_{m{\bf q}}
=
|\phi_{m{\bf q},x_+}|^2
+
|\phi_{m{\bf q},x_-}|^2,
\qquad
{\cal P}^{z}_{m{\bf q}}
=
1-{\cal P}^{x}_{m{\bf q}} .
\label{eq:orbital-projection-numerics}
\end{equation}
For each pocket, the average orbital weight is evaluated as
\begin{equation}
\overline{\cal P}^{r}_{u}
=
\frac{
\sum_{m\in u}\int_{\rm BZ} d^2q\,
{\cal P}^{r}_{m{\bf q}}
\delta_\eta(E_{m{\bf q}})
}{
\sum_{m\in u}\int_{\rm BZ} d^2q\,
\delta_\eta(E_{m{\bf q}})
}.
\label{eq:pocket-averaged-orbital-weight}
\end{equation}
The orbital-selective Hall contribution in Table~\ref{tab:pocket-rh} is
\begin{equation}
R_H^r
=
\sum_u
\overline{\cal P}^{r}_{u}R_H^u.
\label{eq:rh-orbital-numerics}
\end{equation}

The numerical integrations use a $2048\times2048$ ${\bf k}$ mesh with $a=3.754$~\AA{} and $c=20.819$~\AA. We take $\eta=5~{\rm meV}$ and $\Delta_{\alpha\beta}=15~{\rm meV}$ for the main numerical results. Tables~\ref{tab:delta-ab-hall} and~\ref{tab:eta-hall} show the dependence of the vacancy-free Hall coefficient on the small inter-pocket coupling $\Delta_{\alpha\beta}$ and the Dirac-delta broadening $\eta$, respectively. The data in the two tables indicate that the chosen $\Delta_{\alpha\beta}$ and $\eta$ are sufficiently small and have a negligible effect on the numerical results.

\begin{center}
\refstepcounter{table}\label{tab:delta-ab-hall}
\small\textbf{TABLE~\thetable.} Dependence of the vacancy-free Hall coefficient on the inter-pocket coupling $\Delta_{\alpha\beta}$, with $\eta=5~{\rm meV}$.
\begin{ruledtabular}
\begin{tabular}{rrrr}
$\Delta_{\alpha\beta}$ (meV) & $R_H$ ($10^{-3}$ cm$^3$/C) & $R_H-R_H(15~{\rm meV})$ ($10^{-3}$ cm$^3$/C) & $[R_H-R_H(15~{\rm meV})]/R_H(15~{\rm meV})$ (\%) \\
\hline
$0$ & $-0.2889$ & $-0.0072$ & $+2.5663$ \\
$5$ & $-0.2896$ & $-0.0079$ & $+2.7984$ \\
$10$ & $-0.2868$ & $-0.0051$ & $+1.8034$ \\
\baselineboxrow{$15$}{$-0.2817$}{$0.0000$}{$0.0000$}
$20$ & $-0.2751$ & $+0.0065$ & $-2.3227$ \\
\end{tabular}
\end{ruledtabular}
\end{center}

\begin{center}
\refstepcounter{table}\label{tab:eta-hall}
\small\textbf{TABLE~\thetable.} Dependence of the vacancy-free Hall coefficient on the Dirac-delta broadening $\eta$, with $\Delta_{\alpha\beta}=15~{\rm meV}$.
\begin{ruledtabular}
\begin{tabular}{rrrr}
$\eta$ (meV) & $R_H$ ($10^{-3}$ cm$^3$/C) & $R_H-R_H(5~{\rm meV})$ ($10^{-3}$ cm$^3$/C) & $[R_H-R_H(5~{\rm meV})]/R_H(5~{\rm meV})$ (\%) \\
\hline
$3$ & $-0.2816$ & $+0.00005$ & $-0.0181$ \\
$4$ & $-0.2817$ & $+0.00003$ & $-0.0103$ \\
\baselineboxrow{$5$}{$-0.2817$}{$0.0000$}{$0.0000$}
$6$ & $-0.2817$ & $-0.00004$ & $+0.0132$ \\
$7$ & $-0.2818$ & $-0.00008$ & $+0.0292$ \\
\end{tabular}
\end{ruledtabular}
\end{center}

\clearpage

\section{Supplemental Material D: Oxygen-vacancy-induced tight-binding parameter changes from DFT calculations}
\label{app:vacancy-tb}
\renewcommand{\theequation}{D.\arabic{equation}}
\renewcommand{\theHequation}{D.\arabic{equation}}
\setcounter{equation}{0}

The DFT calculations used VASP with the PBE-GGA functional, a plane-wave cutoff of $500~{\rm eV}$, fixed in-plane lattice constants $a=b=3.7544~\text{\AA}$, and a $40~\text{\AA}$ vacuum layer along the $c$ direction. Single-vacancy structures were modeled in $4\times4$ supercells with a $3\times3\times1$ $k$-mesh; after removing an oxygen atom, all internal coordinates were relaxed until the residual forces were smaller than $0.1~{\rm eV}/\text{\AA}$. The DFT+$U$ correction was included in the Dudarev form with $U=4~{\rm eV}$, and the vacancy-free and single-vacancy structures were treated using the same DFT-Wannier workflow with initial projections onto the Ni $d_{x^2-y^2}$ and $d_{z^2}$ orbitals.

The tight-binding parameters used to construct the local vacancy potential are extracted from these DFT calculations. Figures~\ref{fig:apical-vacancy-schematic} and~\ref{fig:inplane-vacancy-schematic} show the corresponding definitions of the tight-binding parameters for inner-apical and in-plane oxygen vacancies. Tables~\ref{tab:apical-vacancy-tb} and~\ref{tab:inplane-vacancy-tb} list the vacancy-free and relaxed values, $p_0$ and $p_{\rm vac}$, for a single vacancy in a $4\times4$ supercell, together with $\Delta p=p_{\rm vac}-p_0$, sorted by decreasing $|\Delta p|$. All parameters are given in eV.

\begin{center}
\refstepcounter{figure}\label{fig:apical-vacancy-schematic}
\includegraphics[width=0.90\textwidth]{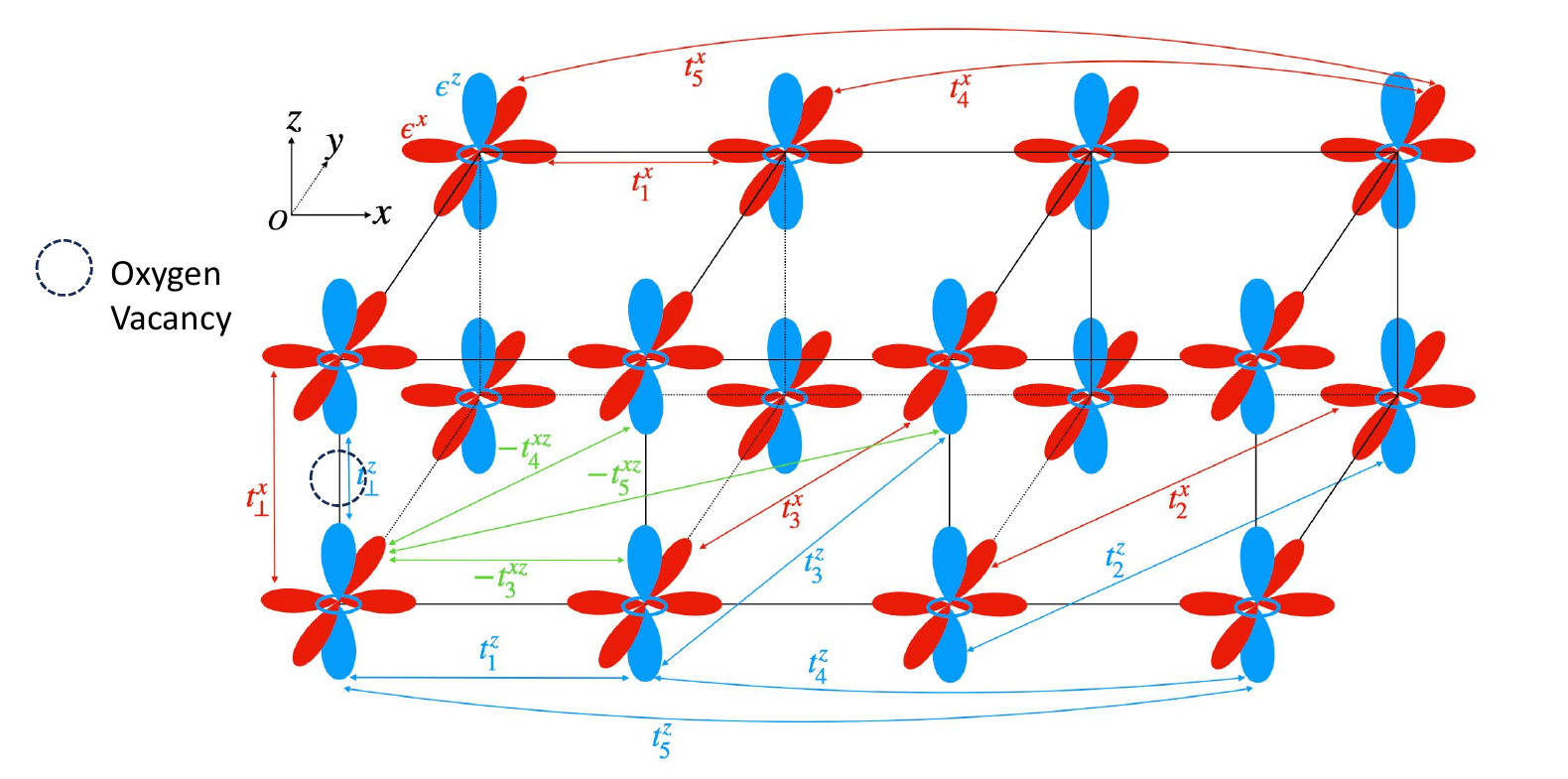}\\[1mm]
\small\textbf{FIG.~\thefigure.} Schematic defining the tight-binding parameters for an inner-apical oxygen vacancy.
\end{center}

\begin{center}
\refstepcounter{table}\label{tab:apical-vacancy-tb}
\scriptsize\textbf{TABLE~\thetable.} Inner-apical oxygen vacancy: vacancy-free and relaxed values of the single-vacancy tight-binding parameters in a $4\times4$ supercell.
\begin{ruledtabular}
\begin{tabular}{lrrr}
Parameter & $p_0$ & $p_{\rm vac}$ & $\Delta p$ \\
\hline
$\epsilon_z$ & $+0.2844$ & $-0.8401$ & $-1.1245$ \\
$\epsilon_x$ & $+0.8107$ & $+1.3530$ & $+0.5423$ \\
$t_\perp^z$ & $-0.5190$ & $-0.0692$ & $+0.4498$ \\
$t_1^z$ & $-0.1290$ & $-0.0486$ & $+0.0804$ \\
$t_2^z$ & $-0.0141$ & $+0.0072$ & $+0.0213$ \\
$t_3^z$ & $+0.0214$ & $+0.0407$ & $+0.0193$ \\
$t_1^x$ & $-0.4646$ & $-0.4459$ & $+0.0187$ \\
$t_4^{xz}$ & $+0.0280$ & $+0.0205$ & $-0.0075$ \\
$t_5^z$ & $-0.0035$ & $+0.0024$ & $+0.0059$ \\
$t_5^{xz}$ & $-0.0006$ & $-0.0039$ & $-0.0033$ \\
$t_3^x$ & $-0.0001$ & $+0.0025$ & $+0.0026$ \\
$t_2^x$ & $+0.0677$ & $+0.0652$ & $-0.0025$ \\
$t_\perp^x$ & $-0.0005$ & $+0.0017$ & $+0.0022$ \\
$t_5^x$ & $-0.0154$ & $-0.0145$ & $+0.0009$ \\
$t_4^x$ & $-0.0559$ & $-0.0551$ & $+0.0008$ \\
$t_3^{xz}$ & $-0.2242$ & $-0.2235$ & $+0.0007$ \\
$t_4^z$ & $-0.0129$ & $-0.0132$ & $-0.0003$ \\
\end{tabular}
\end{ruledtabular}
\end{center}

\begin{center}
\refstepcounter{figure}\label{fig:inplane-vacancy-schematic}
\includegraphics[width=0.92\textwidth]{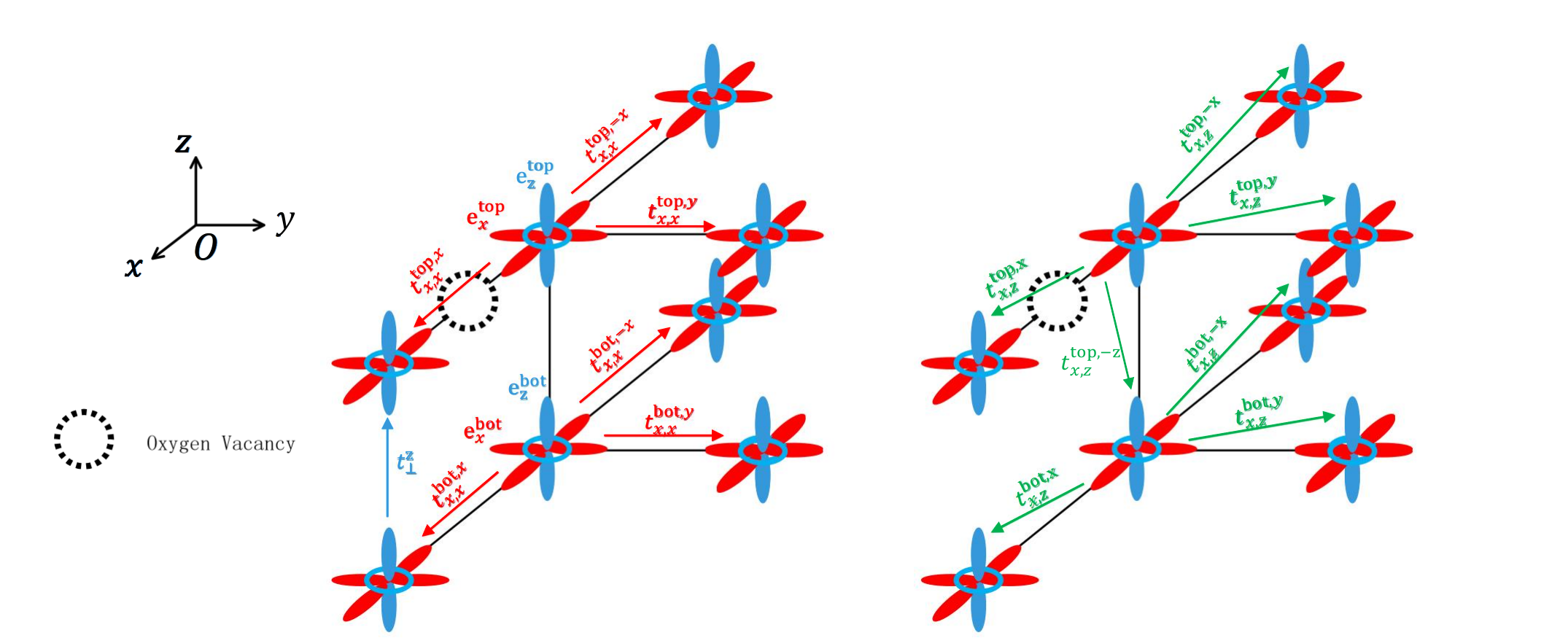}\\[1mm]
\small\textbf{FIG.~\thefigure.} Schematic defining the tight-binding parameters for an in-plane oxygen vacancy.
\end{center}

\begin{center}
\refstepcounter{table}\label{tab:inplane-vacancy-tb}
\scriptsize\textbf{TABLE~\thetable.} In-plane oxygen vacancy: vacancy-free and relaxed values of the single-vacancy tight-binding parameters in a $4\times4$ supercell.
\begin{ruledtabular}
\begin{tabular}{lrrr}
Parameter & $p_0$ & $p_{\rm vac}$ & $\Delta p$ \\
\hline
$\epsilon_x^{\rm top}$ & $+0.8096$ & $-0.5312$ & $-1.3408$ \\
$\epsilon_z^{\rm top}$ & $+0.2840$ & $+0.5444$ & $+0.2604$ \\
$t_{x,z}^{\rm bot,-x}$ & $+0.2242$ & $-0.0362$ & $-0.2604$ \\
$t_{x,z}^{\rm bot,y}$ & $-0.2242$ & $+0.0296$ & $+0.2538$ \\
$t_{x,z}^{\rm bot,x}$ & $+0.2242$ & $-0.0229$ & $-0.2471$ \\
$t_{x,x}^{\rm top,y}$ & $-0.4646$ & $-0.2259$ & $+0.2387$ \\
$\epsilon_z^{\rm bot}$ & $+0.2845$ & $+0.1055$ & $-0.1790$ \\
$t_{x,z}^{\rm top,-z}$ & $+0.0000$ & $-0.1455$ & $-0.1455$ \\
$t_\perp^z$ & $-0.5190$ & $-0.4092$ & $+0.1098$ \\
$t_{x,z}^{\rm top,x}$ & $+0.2242$ & $+0.1255$ & $-0.0987$ \\
$t_{x,x}^{\rm top,-x}$ & $-0.4646$ & $-0.3820$ & $+0.0826$ \\
$t_{x,z}^{\rm top,y}$ & $-0.2242$ & $-0.1454$ & $+0.0788$ \\
$t_{x,x}^{\rm top,x}$ & $-0.4646$ & $-0.3892$ & $+0.0754$ \\
$t_{x,z}^{\rm top,-x}$ & $+0.2242$ & $+0.1935$ & $-0.0307$ \\
$\epsilon_x^{\rm bot}$ & $+0.8108$ & $+0.8270$ & $+0.0162$ \\
$t_{x,x}^{\rm bot,-x}$ & $-0.4646$ & $-0.4555$ & $+0.0091$ \\
$t_{x,x}^{\rm bot,x}$ & $-0.4646$ & $-0.4600$ & $+0.0046$ \\
$t_{x,x}^{\rm bot,y}$ & $-0.4646$ & $-0.4638$ & $+0.0008$ \\
\end{tabular}
\end{ruledtabular}
\end{center}

\section{Supplemental Material E: Transport relaxation time calculation from T-matrix and semiclassical Boltzmann transport}
\renewcommand{\theequation}{E.\arabic{equation}}
\renewcommand{\theHequation}{E.\arabic{equation}}
\setcounter{equation}{0}
\label{app:tmatrix-transport}
We summarize the derivation of the transport relaxation time from the $T$ matrix and the Boltzmann equation. In the vacancy-free system, the retarded Green's function in the (anti-)bonding basis is
\begin{equation}
G_0^R({\bf k},\omega)
=
\left[
(\omega+i0^+)\hat I-H_{\rm QP}({\bf k})
\right]^{-1},
\label{eq:g0-tmatrix}
\end{equation}
where $H_{\rm QP}({\bf k})$ is defined in Eq.~(\ref{eq:hqp}). A single oxygen vacancy generates a perturbation
\begin{equation}
H_{\rm vac}
=
\sum_{{\bf R}{\bf R}'}
\sum_{\lambda\lambda'}
\sum_{\sigma}
\Delta V_{\lambda\lambda'}({\bf R},{\bf R}')
c_{{\bf R}\lambda\sigma}^{\dagger}c_{{\bf R}'\lambda'\sigma},
\label{eq:vac-realspace-potential}
\end{equation}
where $\lambda,\lambda'=A_x,A_z,B_x,B_z$ label the layer-orbital degrees of freedom. The matrix $\Delta V_{\lambda\lambda'}({\bf R},{\bf R}')$ contains the onsite and hopping changes listed in Supplemental Material D. Before constructing the $T$ matrix, this real-space potential is rotated to the same (anti-)bonding orbital basis as $H_{\rm QP}$. We denote the corresponding orbital index by $\zeta=z_+,x_+,x_-,z_-$ and write the rotated potential as $V_{\zeta\zeta'}({\bf R},{\bf R}')$. In this basis, the vacancy perturbation is
\begin{equation}
H_{\rm vac}
=
\frac{1}{N}
\sum_{{\bf k}{\bf k}'}
\sum_{\zeta\zeta'}
\sum_{\sigma}
V_{\zeta\zeta'}({\bf k},{\bf k}')
d_{{\bf k}\zeta\sigma}^{\dagger}d_{{\bf k}'\zeta'\sigma},
\label{eq:vac-momentum-potential}
\end{equation}
where $d_{{\bf k}\zeta\sigma}$ annihilates an electron with spin $\sigma$ in the (anti-)bonding orbital $\zeta$. The vacancy potential is spin independent, so the same $T$ matrix applies in each spin sector. The momentum-space potential is
\begin{equation}
V_{\zeta\zeta'}({\bf k},{\bf k}')
=
\sum_{{\bf R}{\bf R}'}
e^{-i{\bf k}\cdot{\bf R}}
V_{\zeta\zeta'}({\bf R},{\bf R}')
e^{i{\bf k}'\cdot{\bf R}'} .
\label{eq:vac-fourier-potential}
\end{equation}
The single-vacancy $T$ matrix sums repeated scattering from the same vacancy,
\begin{equation}
T^R(\omega)
=
V
+
V\,g_0^R(\omega)\,T^R(\omega),
\label{eq:tmatrix-ls}
\end{equation}
where the vacancy cluster denotes the finite set of real-space sites directly connected by the nonzero vacancy-induced onsite and hopping changes in $\Delta V$. The vacancy-cluster basis is $|{\bf R},\zeta\rangle$, with composite index $a=({\bf R},\zeta)$, so that $V$ and $T^R$ are finite matrices in this basis. In this notation, $V_{ab}$ is the rotated real-space vacancy potential and $T_{ab}^R(\omega)$ is the corresponding multiple-scattering matrix in the same cluster subspace. The projected vacancy-free propagator is the pristine Green's function restricted to this cluster basis,
\begin{equation}
g_{0,{\bf R}\zeta;{\bf R}'\zeta'}^R(\omega)
=
\frac{1}{N}
\sum_{\bf k}
\left[G_0^R({\bf k},\omega)\right]_{\zeta\zeta'}
e^{i{\bf k}\cdot({\bf R}-{\bf R}')}.
\label{eq:local-g0}
\end{equation}
Solving Eq.~(\ref{eq:tmatrix-ls}) gives
\begin{equation}
T^R(\omega)
=
\left[
\hat I-V\,g_0^R(\omega)
\right]^{-1}
V .
\label{eq:tmatrix-solution}
\end{equation}
Its band-basis matrix element between quasiparticle states $|m{\bf k}\rangle$ and $|m'{\bf k}'\rangle$ is
\begin{equation}
T_{m'm}^R({\bf k}',{\bf k};\omega)
=
\Phi_{m'{\bf k}'}^{\dagger}
T^R(\omega)
\Phi_{m{\bf k}},
\label{eq:band-tmatrix}
\end{equation}
where
\begin{equation}
\left[\Phi_{m{\bf k}}\right]_{{\bf R}\zeta}
=
e^{i{\bf k}\cdot{\bf R}}
\phi_{m{\bf k},\zeta}
\label{eq:cluster-bloch-state}
\end{equation}
is the representation of the Bloch eigenstate on the finite real-space vacancy cluster. Thus $\Phi_{m{\bf k}}$ contains the same orbital eigenvector $\phi_{m{\bf k}}$ as in the main text, supplemented only by the Bloch phase factors needed to contract with the cluster matrices $T^R$ and $V$.

For a dilute oxygen-vacancy concentration $n_{\rm vac}$, Fermi's golden rule gives the elastic transition probability
\begin{equation}
W_{m{\bf k}\rightarrow m'{\bf k}'}
=
\frac{2\pi}{\hbar}
n_{\rm vac}
\left|
T_{m'm}^R({\bf k}',{\bf k};E_{m{\bf k}})
\right|^2
\delta
\left(
E_{m{\bf k}}-E_{m'{\bf k}'}
\right).
\label{eq:transition-probability}
\end{equation}
Here $m$ and $m'$ label quasiparticle bands, and $\delta$ denotes the Dirac delta function ensuring elastic energy conservation.

The transition probability enters the Boltzmann collision integral, rather than the conductivity formula directly. For the distribution function $f_{m{\bf k}}$, the impurity collision integral is
\begin{equation}
{\cal I}_{m{\bf k}}[f]=\sum_{m'}\int_{\rm BZ}\frac{d^2k'}{(2\pi)^2}\bigl[W_{m'{\bf k}'\rightarrow m{\bf k}}f_{m'{\bf k}'}(1-f_{m{\bf k}})-W_{m{\bf k}\rightarrow m'{\bf k}'}f_{m{\bf k}}(1-f_{m'{\bf k}'})\bigr] .
\label{eq:collision-integral}
\end{equation}
In a uniform electric field, write the nonequilibrium distribution as
\begin{equation}
f_{m{\bf k}}
=
f_0(E_{m{\bf k}})
-
e
\left(-\frac{\partial f_0}{\partial E}\right)
{\bf E}\cdot{\boldsymbol\Lambda}_{m{\bf k}},
\label{eq:distribution-lambda}
\end{equation}
where ${\boldsymbol\Lambda}_{m{\bf k}}$ is the vector mean free path. The steady homogeneous Boltzmann equation in zero magnetic field is
\begin{equation}
-e\,{\bf E}\cdot\frac{1}{\hbar}\nabla_{\bf k}f_{m{\bf k}}
=
{\cal I}_{m{\bf k}}[f],
\label{eq:steady-boltzmann-tmatrix}
\end{equation}
where \(e>0\). The band velocity is
\begin{equation}
{\bf v}_{m{\bf k}}
=
\frac{1}{\hbar}\nabla_{\bf k}E_{m{\bf k}} .
\label{eq:band-velocity-tmatrix}
\end{equation}
To linear order in the electric field, the left-hand side of Eq.~(\ref{eq:steady-boltzmann-tmatrix}) becomes
\begin{equation}
-e\,{\bf E}\cdot\frac{1}{\hbar}\nabla_{\bf k}f_0(E_{m{\bf k}})
=
e\left(-\frac{\partial f_0}{\partial E}\right)
{\bf E}\cdot{\bf v}_{m{\bf k}} .
\label{eq:driving-term-tmatrix}
\end{equation}
For the collision integral, elastic scattering imposes \(E_{m{\bf k}}=E_{m'{\bf k}'}\), so the equilibrium occupation and its energy derivative are the same for the incoming and outgoing states. Together with detailed balance, \(W_{m'{\bf k}'\rightarrow m{\bf k}}=W_{m{\bf k}\rightarrow m'{\bf k}'}\), Eq.~(\ref{eq:collision-integral}) gives
\begin{equation}
{\cal I}_{m{\bf k}}^{(1)}
=
e\left(-\frac{\partial f_0}{\partial E}\right)
{\bf E}\cdot
\sum_{m'}
\int_{\rm BZ}\frac{d^2k'}{(2\pi)^2}
W_{m{\bf k}\rightarrow m'{\bf k}'}
\left(
{\boldsymbol\Lambda}_{m{\bf k}}
-
{\boldsymbol\Lambda}_{m'{\bf k}'}
\right).
\label{eq:linearized-collision-integral}
\end{equation}
Equating the coefficients of \({\bf E}\) on the two sides gives
\begin{equation}
{\bf v}_{m{\bf k}}
=
\sum_{m'}
\int_{\rm BZ}\frac{d^2k'}{(2\pi)^2}
W_{m{\bf k}\rightarrow m'{\bf k}'}
\left(
{\boldsymbol\Lambda}_{m{\bf k}}
-
{\boldsymbol\Lambda}_{m'{\bf k}'}
\right).
\label{eq:linearized-boltzmann-tmatrix}
\end{equation}
Here the first term is the out-scattering contribution proportional to the vector mean free path of the incoming state, while the second term is the in-scattering, or vertex-correction, contribution from the scattered states.
The coefficient multiplying ${\boldsymbol\Lambda}_{m{\bf k}}$ in Eq.~(\ref{eq:linearized-boltzmann-tmatrix}) defines the single-particle elastic rate,
\begin{equation}
\frac{1}{\tau_{m{\bf k}}^{\rm sp}}
=
\sum_{m'}
\int_{\rm BZ}\frac{d^2k'}{(2\pi)^2}
W_{m{\bf k}\rightarrow m'{\bf k}'},
\label{eq:single-particle-rate-tmatrix}
\end{equation}
Using this definition, Eq.~(\ref{eq:linearized-boltzmann-tmatrix}) can be rewritten as
\begin{equation}
{\bf v}_{m{\bf k}}
=
\frac{{\boldsymbol\Lambda}_{m{\bf k}}}{\tau_{m{\bf k}}^{\rm sp}}
-
\sum_{m'}
\int_{\rm BZ}\frac{d^2k'}{(2\pi)^2}
W_{m{\bf k}\rightarrow m'{\bf k}'}
{\boldsymbol\Lambda}_{m'{\bf k}'}.
\label{eq:lambda-rearranged-tmatrix}
\end{equation}
Solving Eq.~(\ref{eq:lambda-rearranged-tmatrix}) for the vector mean free path yields
\begin{equation}
{\boldsymbol\Lambda}_{m{\bf k}}
=
\tau_{m{\bf k}}^{\rm sp}
\left[
{\bf v}_{m{\bf k}}
+
\sum_{m'}
\int_{\rm BZ}\frac{d^2k'}{(2\pi)^2}
W_{m{\bf k}\rightarrow m'{\bf k}'}
{\boldsymbol\Lambda}_{m'{\bf k}'}
\right].
\label{eq:lambda-integral-equation}
\end{equation}
Equation~(\ref{eq:lambda-integral-equation}) is the vertex-corrected Boltzmann equation generated by the $T$-matrix scattering probability. The second term in the square brackets is the transport vertex correction. It shows why the quasiparticle lifetime obtained by summing all scattering events is not generally the same as the transport relaxation time.

For the relaxation-time estimate used in the main text, the vector mean free path is approximated by ${\boldsymbol\Lambda}_{m{\bf k}}=\tau_{m{\bf k}}{\bf v}_{m{\bf k}}$. In the scalar transport-time approximation, the magnitude of the vector mean free path on the elastic scattering shell is represented by the incoming state, while its direction follows the band velocity. Projecting Eq.~(\ref{eq:linearized-boltzmann-tmatrix}) onto $\hat{\bf v}_{m{\bf k}}$ then gives
\begin{equation}
\left|{\bf v}_{m{\bf k}}\right|
=
\hat{\bf v}_{m{\bf k}}\cdot
\sum_{m'}
\int_{\rm BZ}\frac{d^2k'}{(2\pi)^2}
W_{m{\bf k}\rightarrow m'{\bf k}'}
\left(
{\boldsymbol\Lambda}_{m{\bf k}}
-
{\boldsymbol\Lambda}_{m'{\bf k}'}
\right).
\label{eq:velocity-projection-tmatrix}
\end{equation}
Using the scalar transport-time approximation in Eq.~(\ref{eq:velocity-projection-tmatrix}) gives
\begin{equation}
\left|{\bf v}_{m{\bf k}}\right|
\simeq
\tau_{m{\bf k}}
\left|{\bf v}_{m{\bf k}}\right|
\sum_{m'}
\int_{\rm BZ}\frac{d^2k'}{(2\pi)^2}
W_{m{\bf k}\rightarrow m'{\bf k}'}
\left(
1-\hat{\bf v}_{m{\bf k}}\cdot\hat{\bf v}_{m'{\bf k}'}
\right).
\label{eq:scalar-transport-approx}
\end{equation}
Therefore the scalar transport relaxation rate is
\begin{equation}
\frac{1}{\tau_{m{\bf k}}}
=
\sum_{m'}
\int_{\rm BZ}\frac{d^2k'}{(2\pi)^2}
W_{m{\bf k}\rightarrow m'{\bf k}'}
\left(
1-\hat{\bf v}_{m{\bf k}}\cdot\hat{\bf v}_{m'{\bf k}'}
\right).
\label{eq:transport-rate-w}
\end{equation}
Substituting the $T$-matrix transition probability in Eq.~(\ref{eq:transition-probability}) into Eq.~(\ref{eq:transport-rate-w}) gives the working expression
\begin{equation}
\begin{aligned}
\frac{1}{\tau_{m{\bf k}}}
&=
\frac{2\pi}{\hbar}n_{\rm vac}
\sum_{m'}
\int_{\rm BZ}\frac{d^2k'}{(2\pi)^2}
\left|
T_{m'm}^R({\bf k}',{\bf k};E_{m{\bf k}})
\right|^2
\left(
1-\hat{\bf v}_{m{\bf k}}\cdot\hat{\bf v}_{m'{\bf k}'}
\right)
\\
&\quad\times
\delta
\left(
E_{m{\bf k}}-E_{m'{\bf k}'}
\right),
\end{aligned}
\label{eq:tmatrix-rate}
\end{equation}
where $\hat{\bf v}_{m{\bf k}}={\bf v}_{m{\bf k}}/|{\bf v}_{m{\bf k}}|$. In the numerical evaluation, the Dirac delta function in Eq.~(\ref{eq:tmatrix-rate}) is implemented as a broadened kernel with $\eta=5~{\rm meV}$. The factor $1-\hat{\bf v}_{m{\bf k}}\cdot\hat{\bf v}_{m'{\bf k}'}$ suppresses forward scattering and weights backscattering more strongly; this is the essential distinction between the single-particle lifetime in Eq.~(\ref{eq:single-particle-rate-tmatrix}) and the transport relaxation time in Eq.~(\ref{eq:tmatrix-rate}). For a mixture of vacancy configurations, the total transport relaxation time is obtained from Matthiessen's rule in Eq.~(\ref{eq:matthiessen-main}).

\clearpage

\section{Supplemental Material F: Additional transport analysis for oxygen-vacancy scattering}
\label{app:tau0-dependence}

We further examine how the Hall response depends on the assumed vacancy-free transport relaxation time $\tau_0$. Figure~\ref{fig:tau0-dependence-appendix} shows the same calculation as Fig.~\ref{fig:hall-vacancy-scattering}, but with $\tau_0=10$, $100$, $200$, and $500~{\rm fs}$. Increasing $\tau_0$ reduces the background scattering rate, so a smaller oxygen-vacancy concentration is needed for oxygen-vacancy scattering to dominate the transport. Consequently, the sign-change concentration $n_{\rm vac}^{\ast}$ shifts to smaller values as $\tau_0$ increases. The qualitative trend, namely that distributions with more in-plane oxygen vacancies drive $R_H$ toward positive values more efficiently, remains unchanged.

\begin{center}
\refstepcounter{figure}\label{fig:tau0-dependence-appendix}
\includegraphics[width=0.86\textwidth]{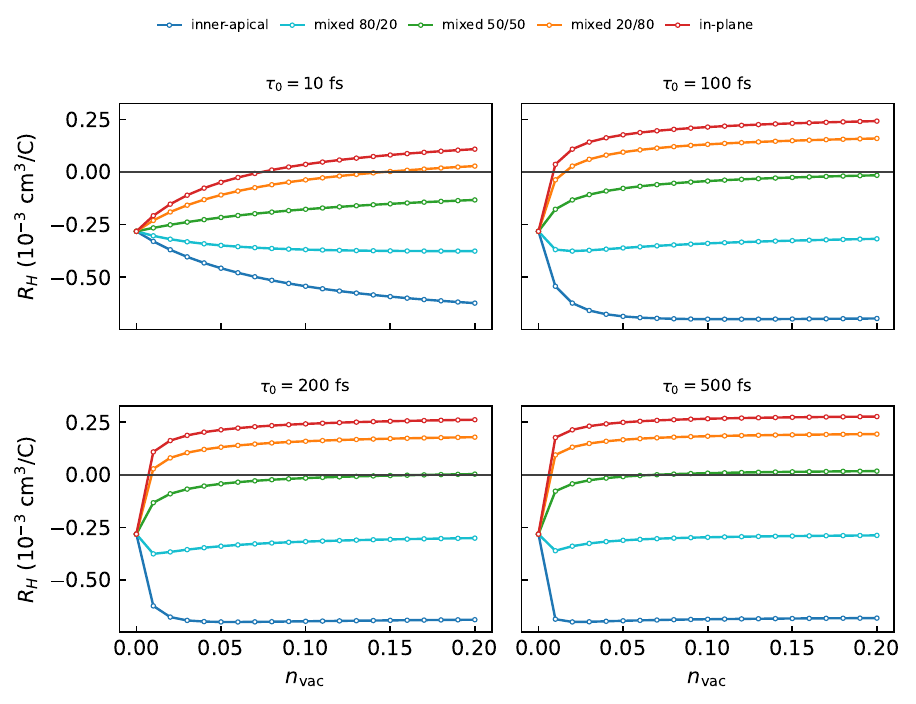}\\[1mm]
\small\textbf{FIG.~\thefigure.} Dependence of $R_H$ on the assumed vacancy-free transport relaxation time $\tau_0$. The four panels show calculations with $\tau_0=10$, $100$, $200$, and $500~{\rm fs}$ for the same five oxygen-vacancy distributions as in Fig.~\ref{fig:hall-vacancy-scattering}.
\end{center}

For the mixed $20\%$ inner-apical and $80\%$ in-plane oxygen-vacancy distribution, Table~\ref{tab:appendix-f-conductivity-decomp} lists the conductivity components at $n_{\rm vac}=0$ and $0.1$. The orbital entries are obtained using the same pocket-averaged orbital projection as in Supplemental Material C.

\begin{center}
\refstepcounter{table}\label{tab:appendix-f-conductivity-decomp}
\small\textbf{TABLE~\thetable.} Pocket-resolved and orbital-selective conductivity components for the mixed $20\%$ inner-apical and $80\%$ in-plane oxygen-vacancy distribution. The longitudinal conductivities are in units of $10^5~\Omega^{-1}{\rm m}^{-1}$, $\sigma_{xy}/B$ is in units of $10^3~\Omega^{-1}{\rm m}^{-1}{\rm T}^{-1}$, and $R_H$ is in units of $10^{-3}~{\rm cm^3/C}$.
\setlength{\tabcolsep}{3.2pt}
\begin{ruledtabular}
\begin{tabular}{lrrrrrrrr}
Contribution & $\sigma_{xx}(0)$ & $\sigma_{xx}(0.1)$ & $\sigma_{yy}(0)$ & $\sigma_{yy}(0.1)$ & $\sigma_{xy}(0)/B$ & $\sigma_{xy}(0.1)/B$ & $R_H(0)$ & $R_H(0.1)$ \\
\hline
pocket $\alpha$ & $8.6837$ & $1.1266$ & $8.6837$ & $1.0582$ & $-6.2811$ & $-0.0941$ & $-1.2845$ & $-0.6678$ \\
pocket $\beta$ & $12.9817$ & $2.4052$ & $12.9817$ & $2.4197$ & $+4.8873$ & $+0.1025$ & $+0.9995$ & $+0.7273$ \\
pocket $\gamma$ & $0.4479$ & $0.2631$ & $0.4479$ & $0.2344$ & $+0.0157$ & $+0.0049$ & $+0.0032$ & $+0.0348$ \\
orbital $d_{x^2-y^2}$ & $10.2712$ & $1.5810$ & $10.2712$ & $1.5402$ & $-2.5148$ & $-0.0276$ & $-0.5143$ & $-0.1958$ \\
orbital $d_{z^2}$ & $11.8422$ & $2.2140$ & $11.8422$ & $2.1721$ & $+1.1368$ & $+0.0409$ & $+0.2325$ & $+0.2901$ \\
sum & $22.1133$ & $3.7950$ & $22.1133$ & $3.7124$ & $-1.3780$ & $+0.0133$ & $-0.2818$ & $+0.0943$ \\
\end{tabular}
\end{ruledtabular}
\end{center}

Figure~\ref{fig:appendix-f-rigid-band-shift} compares the fixed-$\mu$ and rigid-band shifted-$\mu$ results. For each $n_{\rm vac}$, the shifted chemical potential is first determined from the rigid-band calculation, and the same calculation as Fig.~\ref{fig:hall-vacancy-scattering} is then repeated at this shifted $\mu$.

\begin{center}
\refstepcounter{figure}\label{fig:appendix-f-rigid-band-shift}
\includegraphics[width=0.58\textwidth]{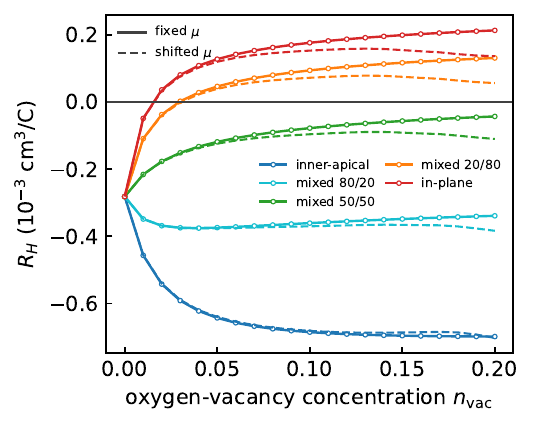}\\[1mm]
\small\textbf{FIG.~\thefigure.} Comparison between the fixed-$\mu$ and rigid-band shifted-$\mu$ results. Solid curves are just Fig.~\ref{fig:hall-vacancy-scattering}; dashed curves are obtained by first determining the rigid-band shifted chemical potential at each $n_{\rm vac}$ and then repeating the same calculation as Fig.~\ref{fig:hall-vacancy-scattering} at that $\mu$.
\end{center}

\end{document}